\documentclass[sigconf,nonacm,prologue,table,xcdraw]{acmart}

\usepackage[utf8]{inputenc}
\usepackage{url}
\usepackage{graphicx}
\usepackage{listings}
\usepackage{epstopdf}
\usepackage{color}
\usepackage{subfigure}
\usepackage{xspace}
\usepackage{latexsym}
\usepackage{wrapfig}
\usepackage{setspace}
\usepackage{blindtext}
\usepackage{amsfonts}
\usepackage{multirow}
\usepackage{pifont}
\usepackage{mathrsfs}
\usepackage{mathtools}
\usepackage{relsize}
\usepackage{lipsum}

\usepackage{graphics}

\usepackage{ulem}
\usepackage{comment}
\usepackage{footnote}
\usepackage{pifont}
\usepackage{makecell}
\usepackage{multirow}
\usepackage{booktabs}
\usepackage{hhline}
\usepackage{adjustbox}

\usepackage{enumitem}

\usepackage{zref}
\usepackage{cleveref}
\crefformat{section}{\S#2#1#3}

\usepackage{algorithm, algpseudocode}
\usepackage[algo2e,ruled,vlined]{algorithm2e}

\newcounter{protocol}
\newenvironment{protocol}[1]
  {\par\addvspace{\topsep}
   \noindent
   \tabularx{\linewidth}{@{} X @{}}
    \hline
    \refstepcounter{protocol}\textbf{Protocol \theprotocol} #1 \\
    \hline}
  { \\
    \hline
   \endtabularx
   \par\addvspace{\topsep}}

\newcommand{\sbline}{\\[.5\normalbaselineskip]}

\newcommand{\proj}{\textsc{$\lambda$FS}\xspace}

\newcommand{\lindexfs}{\textsc{$\lambda$IndexFS}\xspace}
\newcommand{\infcache}{\textsc{InfiniCache}\xspace}

\newif{\ifSubmit}
\newif{\ifFinal}
\newif{\ifDraft}

\Finaltrue 
\Submittrue 

\ifSubmit
\newcommand{\yuec}[1]{}
\newcommand{\bencomment}[1]{}
\newcommand{\jycomment}[1]{}
\newcommand{\maicomment}[1]{}
\newcommand{\rzcomment}[1]{}
\else
\newcommand{\added}[1]{\noindent\textcolor{red}{#1}}
\newcommand{\yuec}[1]{\noindent\textcolor{red}{Yue: #1}}
\newcommand{\jycomment}[1]{\noindent\textcolor{orange}{\bf Jingyuan: #1}}
\newcommand{\bencomment}[1]{\noindent\textcolor{blue}{\bf Ben: #1}}
\newcommand{\maicomment}[1]{\textcolor{brown}{\textbf{Mai: #1}}}
\newcommand{\rzcomment}[1]{\textcolor{teal}{\textbf{Lukas: #1}}}
\fi

\ifDraft
\newcommand{\addcomment}[1]{\textcolor{red}{#1}}
\newcommand{\diffcomment}[2]{\textcolor{orange}{#1}}
\newcommand{\delcomment}[1]{\textcolor{orange}{\sout{#1}}}
\else
\newcommand{\addcomment}[1]{#1}
\newcommand{\diffcomment}[2]{#1}
\newcommand{\delcomment}[1]{}
\fi

\begin{document}

\title{{\proj}: A Scalable and Elastic Distributed File System Metadata Service using Serverless Functions}
\subtitle{This preprint will be published at ACM ASPLOS'24}

\author{Benjamin Carver}
\affiliation{%
  \institution{{\it George Mason University}}
  \streetaddress{4400 University Dr}
   \city{Fairfax}
   \state{VA}
   \country{USA}
}
\email{bcarver2@gmu.edu}
\author{Runzhou Han}
\affiliation{%
  \institution{{\it Iowa State University}}
  \streetaddress{4400 University Dr}
   \city{Ames}
   \state{IA}
   \country{USA}
}
\email{hanrz@iastate.edu}
\author{Jingyuan Zhang}
\affiliation{%
  \institution{{\it George Mason University}}
  \streetaddress{4400 University Dr}
   \city{Fairfax}
   \state{VA}
   \country{USA}
}
\email{jzhang33@gmu.edu}
\author{Mai Zheng}
\affiliation{%
  \institution{{\it Iowa State University}}
  \streetaddress{4400 University Dr}
   \city{Ames}
   \state{IA}
   \country{USA}
}
\email{mai@iastate.edu}
\author{Yue Cheng}
\affiliation{%
  \institution{{\it University of Virginia}}
   \city{Charlottesville}
   \state{VA}
   \country{USA}
}
\email{mrz7dp@virginia.edu}
\authornote{Corresponding author}

\settopmatter{printfolios=true}




\begin{abstract}

The metadata service (MDS) sits on the critical path for \addcomment{distributed} file system \addcomment{(DFS)} operations, and \diffcomment{therefore}{as such} it is key to the overall performance of a large-scale \diffcomment{DFS}{}. 
Common ``serverful'' MDS architectures, such as a single server or cluster of servers, have a significant shortcoming: either they are not scalable, or they make it difficult to achieve an optimal balance of performance, resource utilization, and cost. 
A modern MDS requires a novel architecture that addresses this shortcoming.

To this end, we design and implement {\proj}, an elastic, high-performance metadata service for large-scale \diffcomment{DFSes}{distributed file systems}. {\proj} scales a DFS metadata cache \addcomment{elastically} on a FaaS (Function-as-a-Service) platform and synthesizes a series of techniques to overcome the obstacles that are encountered when building large, stateful, and \addcomment{performance-sensitive} applications on FaaS platforms.  
{\proj} takes full advantage of the unique benefits offered by FaaS---elastic scaling and massive parallelism---to realize a highly-optimized metadata service capable of sustaining up to 
$4.13\times$ higher throughput, $90.40\%$ lower latency, $85.99\%$ lower cost, \addcomment{$3.33\times$ better performance-per-cost}, and better resource utilization and efficiency than a state-of-the-art DFS for an industrial workload. 
\end{abstract}

\date{}
\maketitle

\section{Introduction}
\label{sec:intro}

Many different fields in computing have enjoyed successes in part due to the availability of large amounts of data~\cite{spark_nsdi12, mapreduce_osdi04, hadoop, tensorflow_osdi16, mxnet, pytorch, pywren_socc17, numpy, funcx}. 
Data-intensive applications~\cite{nsf_cdse, nsf_data_science_report} in these fields are characterized by varied, heterogeneous I/O patterns in which I/O bottlenecks are not uncommon~\cite{springfs_fast14, nfs_wl_atc08, hadoop_vldb12}. 
Large-scale, distributed file systems (DFSes), such as Google File System (GFS)~\cite{gfs_sosp03} and Hadoop Distributed File System (HDFS)~\cite{hdfs_msst10},
are commonly used by these data-intensive applications. 
DFSes often use an architecture that \textit{decouples} metadata management from file I/O~\cite{gfs_sosp03, hdfs_msst10, lustre}. DFS metadata tracks global file system namespace information, including hierarchical directories and file names. These DFSes use a centralized metadata management component called a metadata service (MDS), which executes file system namespace operations, such as file {\small\texttt{open}}, {\small\texttt{close}}, and {\small\texttt{mv}}.  
In a DFS, client applications acquire a file's permission and location information from the MDS before accessing the file's contents. 
Therefore, the performance of the MDS is key to the overall efficiency of a DFS. 

Scaling the performance of an MDS is challenging.
Using a dedicated server (e.g., GFS~\cite{gfs_sosp03}, HDFS~\cite{hdfs_msst10}) to host the MDS is not scalable and may suffer from poor performance during highly dynamic workloads. 

Researchers have proposed various ways to overcome the scalability challenges of DFS MDSes. IndexFS~\cite{indexfs_sc14} is a middleware that offloads metadata storage and processing to a scaled-out, table-based, key-value store cluster that is co-located with the data storage cluster. \textsc{InfiniFS}~\cite{InfiniFSEfficientMetadata} uses the same scaled-out cluster architecture as IndexFS but with deep optimizations along the metadata processing path. HopsFS~\cite{hopsfs_fast17}, built on HDFS, 
further decouples metadata request handling and metadata storage: it offloads the metadata storage to a distributed, sharded, in-memory database (MySQL NDB Cluster~\cite{ndb}) and utilizes a cluster of \emph{stateless} NameNodes (metadata servers in HDFS terminology) to scale DB query handling. 

While these systems offer scalable MDS solutions with different tradeoffs, they have a common issue: they lack elasticity support at the MDS level. IndexFS and \textsc{InfiniFS} 
employ a fixed cluster of metadata servers and use client-side metadata caching extensively for performance improvement. HopsFS provides no metadata caching on the stateless NameNode side and uses the distributed NameNodes only for handling and scaling client requests. Therefore, HopsFS' performance is capped by the capacity of the backend NDB cluster. All three of these systems require explicit server management and a large amount of server resources to be reserved to host the MDS cluster. As reported in \cite{hopsfs_fast17}, HopsFS requires as many as 60 NameNodes and 12 NDB servers in order to significantly outperform vanilla HDFS, the latter of which typically uses a small cluster of NameNodes for high availability but not performance. Worse, under low load conditions, the scaled-out MDS cluster suffers from low resource utilization.

Serverless computing or Function-as-a-Service (FaaS) has emerged as a new cloud computing model~\cite{aws_lambda, berkeley_serverless_techreport}. FaaS enables developers to break traditionally monolithic, server-based applications into finer-grained serverless \addcomment{(or cloud)} functions, thereby providing a new way of building and scaling applications and services. Developers are tasked with providing the function logic while the FaaS provider is responsible for the notoriously tedious tasks of provisioning, scaling, and managing backend servers that host the serverless functions~\cite{gray_stop}. 


We find that serverless functions provide an appealing
environment in which to host and scale the metadata management component of a large-scale DFS. 
Using serverless functions provides several key advantages. 
\diffcomment{First, serverless functions have CPU and memory resources that are elastically scaled out and in with the functions. This enables the construction of an elastic MDS that can achieve optimal performance by dynamically adapting the amount of resources as the workload shifts. 
Second, the elasticity offered by FaaS can greatly improve cost-efficiency and resource utilization as resources are allocated/deallocated in an on-demand manner and used more efficiently.
Third, the auto-scaling property also alleviates the need for tedious server management. 
}

The aforementioned challenges pertaining to MDS efficiency and the emergence of serverless computing together raise a research question: \emph{Can we use serverless functions in a novel way to build a high-performance, cost-efficient, elastic, and resource-efficient MDS?}


To answer this question, we present {\proj},
the first serverless-function-based, elastic MDS for large-scale DFSes. 
In a nutshell, {\proj} features a novel MDS architecture that combines an elastically-scalable, FaaS-based metadata cache with a persistent, strongly-consistent metadata store. 
\addcomment{To minimize network overhead, {\proj} uses the collective memory of a dynamic fleet of serverless functions for metadata caching.
However, simply implementing a metadata caching layer is insufficient. {\proj} further enables elastic and massively-parallel metadata caching by taking advantage of the auto-scaling offered by FaaS. Not only does this elasticity improve metadata query performance, but it also enables high resource efficiency and low cost. 
Moreover,} 
{\proj} effectively decouples the management of metadata caching (and thus, metadata request processing) and metadata storage so that compute and storage can scale independently. This fully-disaggregated architecture is driven by the observations that real-world MDS workloads are bursty~\cite{hadoop_vldb12, alibaba_workload_socc19, igen_cloud16} and that it is often difficult to manually determine the right MDS deployment scale offline~\cite{hdfs_scalability_login10, linkedin_hdfs}.

Building an elastic serverless MDS for large-scale DFSes requires addressing two sets of unique challenges:
\begin{itemize}[noitemsep,leftmargin=*]

\item First, FaaS platforms have a series of constraints and limitations that make it challenging to support data-intensive, stateful applications efficiently:
(1)~individual serverless functions have limited CPU, memory, and network resources, and thus offer limited data processing, storage, and transfer capacity. 
(2)~Serverless functions occasionally suffer from long cold start times and execution timeouts.
(3)~The typical method to communicate with serverless functions is via HTTP requests, but this can be slow. 
As such, naively porting \diffcomment{the}{a} stateful MDS of a large-scale DFS to a serverless platform leads to poor performance.

\item Second, while FaaS platforms offer auto-scaling and elasticity, a careful, holistic MDS redesign is required to fully utilize these benefits:
(1)~Performance-sensitive systems such as \diffcomment{MDSes}{MDS} require careful treatment to balance the performance and auto-scaling tradeoff.
(2)~\diffcomment{Partitioning the file system}{Considering how to partition the} namespace across a dynamic fleet of serverless functions introduces interesting tradeoffs in a FaaS environment.
(3)~The lack of addressability of serverless functions means that a metadata entry may be stored on multiple functions, therefore introducing metadata consistency issues.


\end{itemize}

{\proj} addresses these challenges by synthesizing several techniques into an end-to-end, serverless MDS system. 
First, 
we find that {\proj} can achieve strong performance using a large number of serverless NameNodes each having relatively small
CPU and memory resources compared to their serverful counterparts. 
{\proj} also leverages a hybrid HTTP-TCP RPC mechanism to 
enable agile, lightweight, and performance-preserving auto-scaling.

Second, {\proj}' FaaS-powered metadata cache consists of $n$ unique serverless function deployments. {\proj} uses path-based hashing of parent directories to partition the file system namespace among the $n$ 
deployments in order to support efficient metadata {\texttt{\small{read}}} operations. Each deployment can automatically scale out to an arbitrary number of concurrently-running function instances that elastically support bursts of metadata requests on hot directories.
{\proj} trades \diffcomment{function-deployment-based}{unlimited, horizontal} auto-scaling \addcomment{(i.e., there is a \textit{fixed} number $n$ of deployments)}
for easy-to-manage, deterministic metadata partitioning. 
Third, {\proj} implements a serverless coherence protocol to provide strong consistency in the presence of an arbitrary number of running ``function instances'' and clients. 

Finally, {\proj} re-implements many DFS maintenance features, such as block reports and DataNode discovery, in a serverless-compatible way by publishing information to the persistent metadata store on a regular interval. 

\addcomment{There are a number of benefits and advantages of using FaaS as the underlying platform for the MDS of a large-scale DFS. Notably, these benefits would be difficult or impossible to realize using a traditional, serverful MDS architecture. First, by using a large number of relatively lightweight serverless functions, overall resource utilization can be improved, which ultimately leads to better performance and cost-efficiency. This cost-efficiency is further enhanced by the pay-per-use pricing model of FaaS, which drastically lowers tenant-side costs without negatively impacting performance. This is quantified using a 
\textit{performance-per-cost} metric in \cref{subsubsec:real-world-cost}.}

\addcomment{In addition to cost-related benefits, the MDS can take advantage of FaaS-based auto-scaling to automatically 
adapt to changes in \diffcomment{request volume}{throughput} without requiring 
management by users or admins. 
When request volume increases, the MDS automatically scale-outs to serve the additional requests. When request volume decreases, the MDS will scale-in, avoiding the problem of low resource utilization and poor cost-efficiency. This completely circumvents the typical under/over-provision problem faced by serverful systems.}


In summary, this paper makes the following contributions:
\begin{itemize}[noitemsep,leftmargin=*]

\item \diffcomment{We identify scalability and performance issues of HopsFS, a state-of-the-art DFS with a scaled-out MDS design.
}{We identify scalability and performance issues of HopsFS, a state-of-the-art DFS that stores fully normalized metadata in a distributed database and scales metadata processing using a cluster of stateless metadata servers.}

\item We explore the design space of serverless metadata management for large-scale DFSes. We identify the key opportunities and challenges of using FaaS for this purpose, and we share insights into how to address these challenges. 

\item We present the design and implementation of {\proj}, a novel metadata service that uses massively-parallel serverless functions to  cache and \addcomment{elastically} scale the metadata workload. To the best of our knowledge, {\proj} is the first 
serverless-function-based MDS for large distributed file systems.

\item We demonstrate {\proj}' generality  by porting {\proj} to
two practical DFSes that have different, scaled-out MDS architectures: HopsFS and 
BeeGFS~\cite{beegfs} enhanced with IndexFS.

\item \diffcomment{We extensively evaluate {\proj} using real-world workloads and microbenchmarks.}{We evaluate {\proj} extensively using real-world workloads and microbenchmarks to better understand the performance implications of using FaaS for large-scale DFSes.}
Our results show that {\proj} achieves up to $4.13\times$
higher throughput, 90.40\% lower latency, and 85.99\% lower cost than HopsFS \addcomment{and $3.33\times$ greater performance-per-cost than HopsFS augmented with a metadata cache} while providing better resource utilization.

\end{itemize}
\section{Background and Motivation}
\label{sec:moti}


%


Different DFS architectures have different tradeoffs, but there is one commonality: all existing solutions use an
architecture that separates metadata management and file data storage management. In this section, we use HDFS and HopsFS, two representative, production-ready DFSes, as examples to illustrate the two generations of MDS architectures used by today's DFSes.
Specifically, HopsFS uses a cluster of scaled-out, stateless metadata servers in front of a scaled-out, strongly-consistent metadata store to support a scalable MDS for the widely used HDFS.
Therefore, HopsFS provides an ideal platform to experiment with and demonstrate the efficacy of {\proj}. This section describes the common limitations of state-of-the-art MDS solutions to further motivate {\proj}.

\begin{figure}[t]
    \centering
    \includegraphics[width=0.47\textwidth]{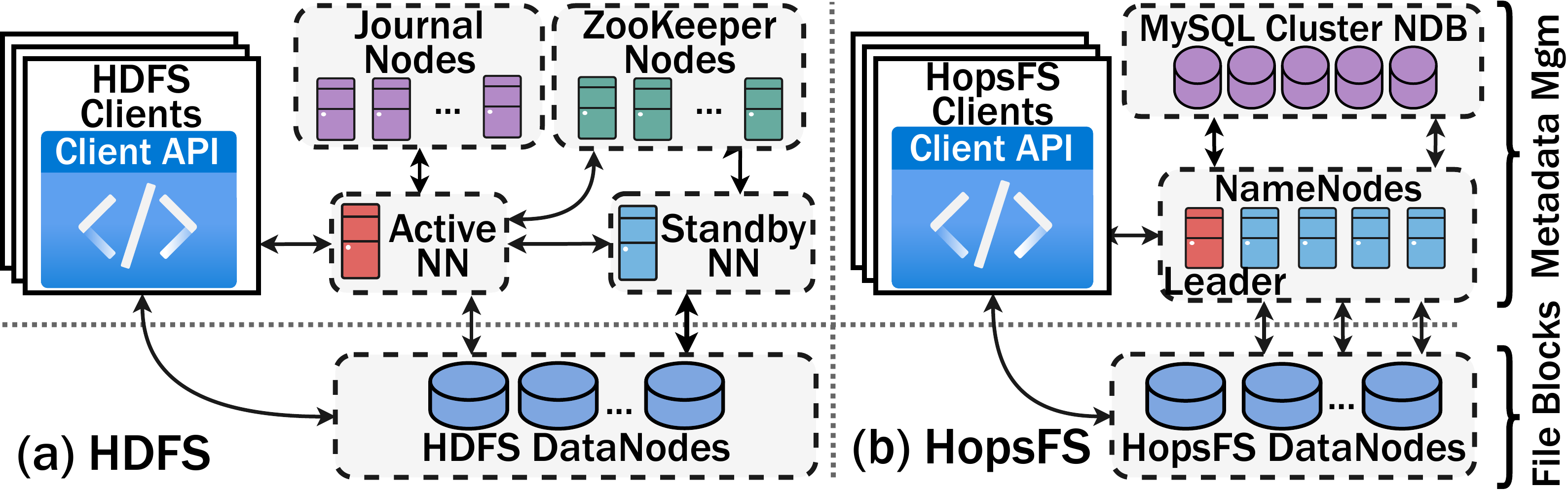}
    \vspace{-10pt}
    \caption{Architecture of HDFS and HopsFS.}
    \label{fig:hdfs-hopsfs-arhcitecture}
    \vspace{-10pt}
\end{figure}

\noindent\textbf{Hadoop Distributed File System.}
The Hadoop Distributed File System (HDFS) is an open-source implementation of GFS \cite{gfs_sosp03} and is widely used in practice~\cite{hdfs_msst10}.
%
HDFS stores metadata in the memory of a Java process referred to as the Active NameNode. This metadata is replicated to a Standby NameNode, which is used for checkpointing and failure recovery. File system operations are executed atomically by the Active NameNode, thereby providing strong consistency for file system metadata. POSIX semantics are relaxed in order to allow for streaming access to the system data. See Figure~\ref{fig:hdfs-hopsfs-arhcitecture}(a).




DataNodes are responsible for storing 
file data. Each DataNode connects to both the Active and Standby NameNodes. A DataNode periodically generates reports that are sent to the NameNode. These reports are used to ensure the NameNode's block map is consistent with the actual data stored in DataNodes. The JournalNodes are used to synchronize state between the Active and Standby NameNodes.  ZooKeeper~\cite{zookeeper_atc10} provides \textit{automatic failover} and \textit{leader election} for the NameNodes. 

\noindent\textbf{HopsFS.}
HopsFS~\cite{hopsfs_fast17} is a distributed file system developed as an extension of HDFS.
HopsFS provides a scaled-out metadata management layer by decoupling the storage and manipulation of metadata. Specifically, HopsFS supports multiple stateless NameNodes. The NameNodes persists the metadata to a pluggable storage backend and collectively serve metadata requests made by clients. HopsFS uses MySQL Cluster NDB \cite{MySQLMySQLReference} for this persistent backend data store. The architecture of HopsFS is shown in Figure~\ref{fig:hdfs-hopsfs-arhcitecture}(b). 



Each NameNode uses a Data Access Layer (DAL) that provides a generic interface to an arbitrary persistent storage backend. This interface is used to manipulate the metadata stored within NDB. 
All file system operations require the resolution of each path component in order to check for permissions and path validity. HopsFS introduces techniques to mitigate the performance impact of path resolutions, such as an ``INode Hint Cache'', which allows clients to cache metadata prefixes locally to reduce the number of round trips required for path resolution from $N$ round trips (for a path of length $N$) to just one single batch query. The cluster of stateless NameNodes cooperates to handle DataNode failures. The NameNodes elect a leader NameNode to perform administrative tasks. 

\noindent\textbf{Limitations of Today's Scaled-Out MDSes.}
For the remainder of the paper, we do not focus on issues that HopsFS has already addressed---{\proj} uses the same decoupled compute-and-storage MDS architecture and uses the same DAL to interface with the persistent metadata store used by HopsFS. Instead, we focus on the scalability and elasticity problems with HopsFS' statically-fixed, stateless NameNode cluster design, which we describe next.

There are several aspects of HopsFS' design that 
hinder HopsFS' MDS efficiency.
First, the use of \textit{stateless} NameNodes necessitates the retrieval of metadata from the persistent metadata store for every single metadata operation. This means that HopsFS' performance is capped by the capacity of the backend NDB cluster. The compute (NameNode) and storage (NDB) resources, though physically decoupled, are essentially \emph{logically-bundled} resources that need to be configured together. Otherwise, 
system performance can rapidly degrade if either of the two layers
becomes a bottleneck.

Second, 
HopsFS and other scaled-out MDS solutions~\cite{InfiniFSEfficientMetadata, indexfs_sc14, ceph_mds_sc04} lack elasticity and require an admin to empirically configure a statically-fixed deployment of compute and storage resources for the serverful MDS cluster. This leads to a choice between resource under-utilization and degraded performance: if the admin provisions compute resources for the peak load of the metadata workload, the system wastes both compute and storage resources; 
if the admin provisions resources for the average demand, then the performance degrades when the load increases beyond the provisioned capacity. 


\noindent\textbf{Terminology.} \label{subsubsec:terminology} Before describing {\proj}' design, it is necessary to define some terminology. First, {\proj}' NameNodes are organized into multiple serverless \textit{function deployments}. Function deployments consist of user-written code to be executed when the serverless function runs, configuration info, and metadata, all of which is registered under a unique name with the FaaS platform. 
The code for a NameNode is written (in Java) as the body of a serverless function. {\proj} registers a configurable number of uniquely named serverless NameNode functions with the FaaS platform. (The bodies of these functions are identical; the names are different.) 

When a user invokes a serverless function defined by a particular deployment, the FaaS platform automatically provisions an \textit{instance} of that function based on the configuration \diffcomment{info}{information} specified when the deployment was registered. Thus, a \textit{function instance} refers to an instantiated, running serverless function. A \textit{NameNode} then refers to the Java application executing within the function instance. When we say that a function instance ``belongs'' to a 
deployment, we mean that the instance is an 
instantiation of the serverless function defined by that deployment. \diffcomment{Only one NameNode can execute within a function instance, so}{Since there can only be one NameNode executing within a function instance,} the two terms are used interchangeably.
\section{{\proj} Design}
\label{sec:design}


{\proj} enables elastic metadata service for large-scale DFSes such as HDFS. {\proj} uses a hybrid, FaaS-optimized RPC mechanism that combines TCP-based RPC and HTTP-based RPC together to enable high throughput and reduced latency (\cref{subsec:design-metadata-requests}). 
\diffcomment{{\proj} uses a serverless-function-based memory caching layer for DFS metadata caching (\cref{subsec:design-metadata-cache}) and features an agile auto-scaling policy to enable elastic and parallel metadata processing at the caching layer (\cref{subsec:design-agile-auto-scaling}).}{{\proj} uses an elastic shared memory pool built on serverless functions to support DFS metadata caching and processing (\cref{subsec:design-metadata-cache}).}
\diffcomment{While caching reduces the number of network hops per request, {\proj}' auto-scaling significantly improves the cache's throughput.}{Elastic, serverless caching reduces the number of network hops per metadata request, which substantially reduces latency.}
{\proj} also introduces a simple coherence protocol to ensure strong consistency of metadata operations within the serverless cache (\cref{subsec:design-consistency-protocol}). 
Note that {\proj} is also usable with other DFSes: as we will show in \cref{sec:implementation}, {\proj} can enhance BeeGFS~\cite{beegfs} as a drop-in replacement for IndexFS.

\subsection{{\proj} Overview}
\label{subsec:design-proj-overview}


\begin{figure}[t]
\begin{center}
\includegraphics[width=0.46\textwidth]{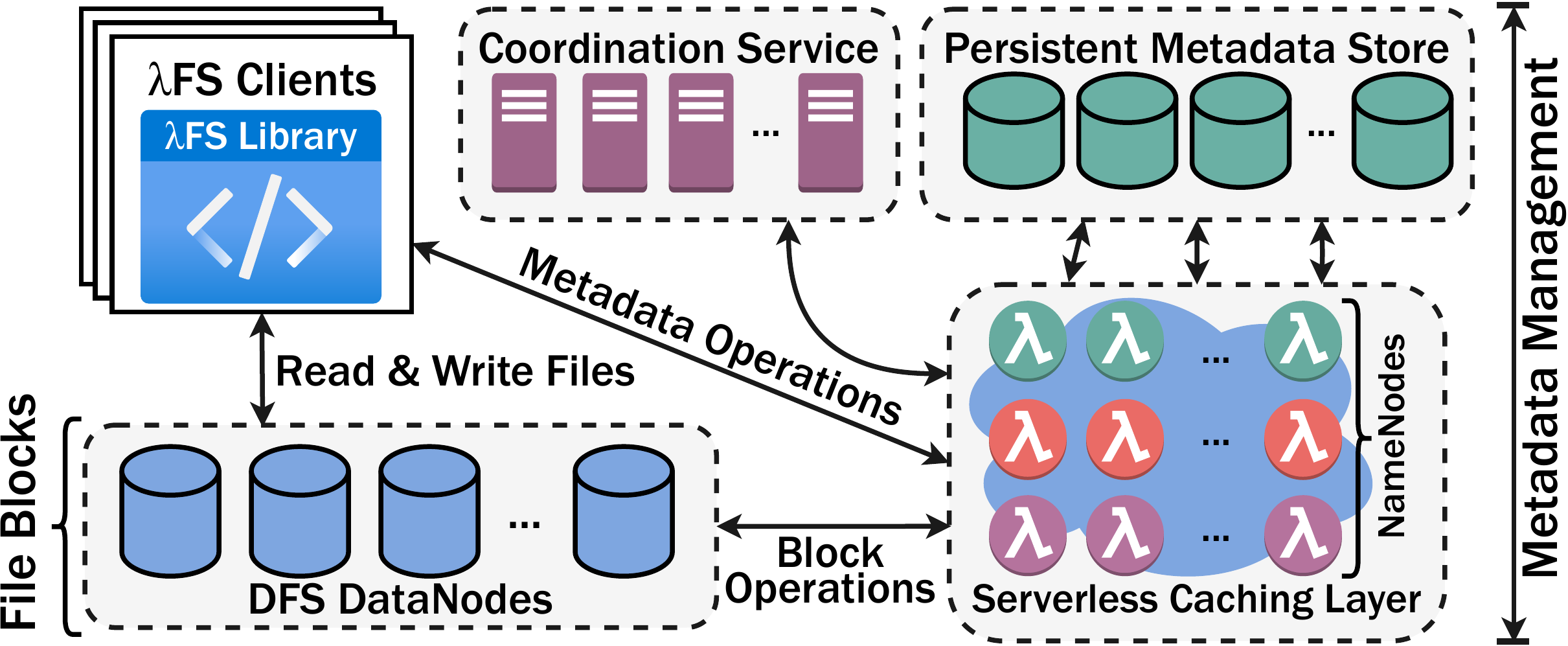}
\vspace{-7pt}
\caption{
The {\proj} architecture.
}
\label{fig:serverless-hopsfs-arhcitecture}
\vspace{-12pt}
\end{center}
\end{figure}

Figure~\ref{fig:serverless-hopsfs-arhcitecture} shows the architecture of {\proj}. 
Clients issue RPC metadata requests to NameNodes, just as clients do in HopsFS. The difference is that each {\proj} NameNode is a Java serverless function 
executing within a container
managed by the serverless platform. RPC metadata requests are initially performed as HTTP invocations directed towards the platform's API gateway. 
The serverless platform routes 
HTTP requests it receives to already-running NameNode instances, if available, or it starts a new 
instance if none are running.

Once a NameNode is up and running, it can establish direct TCP connections back to clients (after first interfacing with clients through HTTP requests). TCP-based RPC requests serve as a lower-latency alternative to HTTP-based RPC requests, as only one network hop (client to NameNode) is required for a TCP RPC. Hybrid TCP and HTTP RPC mechanisms are discussed in \cref{subsec:design-metadata-requests}.

Another key difference between {\proj} and HopsFS is that the serverless NameNodes in {\proj}: 
\diffcomment{are (1)~\emph{not} stateless, and (2)~are \emph{elastic}.}{are \emph{elastic} and are \textit{not} stateless.} 
This allows the dynamic cluster of serverless NameNodes to collectively form an \textit{elastic metadata caching layer}. When a NameNode 
receives a metadata request, it checks whether the requested \diffcomment{metadata was retained from a previous function invocation}{metadata object is a retained object from a previous invocation of the function}. The retained \diffcomment{metadata}{objects} thus form a cache. \addcomment{The caching system is discussed further in \cref{subsec:design-metadata-cache}.}

\diffcomment{To support \emph{elastic caching},}{More specifically,} {\proj}' NameNodes
are organized into $n$ individual \textit{function deployments}.
We partition the namespace among the function deployments by consistently hashing on the parent directory path of each file/directory. For example, we may hash the file ``{\small{\texttt{/dir/note.pdf}}}'' to the deployment named ``{\small{\texttt{NameNode5}}}''. In this case, the client would issue an HTTP RPC for the deployment ``{\small{\texttt{NameNode5}}}'', or issue a TCP RPC to an already-running function instance of the {\small{\texttt{NameNode5}}} deployment. If the NameNode that serves this request already has the target metadata cached \diffcomment{locally}{in memory}, then a network hop to the persistent metadata store is avoided, resulting in lower latency for the request. Individual function deployments automatically scale out
in response to the sudden increase of metadata requests. \addcomment{Our agile auto-scaling policy is described in \cref{subsec:design-agile-auto-scaling}.}

{\proj} uses a pluggable ``Coordinator'' service for tracking NameNode liveness and coordinating NameNodes during write operations. {\proj} currently supports both ZooKeeper and MySQL Cluster NDB. The Coordinator is used in the coherence protocol described in~\cref{subsec:design-consistency-protocol}.

{\proj}' design capitalizes on the unique benefits of serverless computing while accounting for the challenges that the platform presents. 
First, {\proj} takes advantage of the 
intra-deployment auto-scaling offered by FaaS to rapidly and transparently scale-out in response to bursts of work. Not only does this enable {\proj} to responsively and elastically adapt to changing workload characteristics in real-time, but it also improves {\proj}' resource efficiency and resource utilization. When system throughput returns to normal levels, {\proj} will transparently scale-in to avoid incurring additional costs. Using traditional, serverful VMs in place of serverless functions would result in significantly reduced elasticity and either wasted resources or poor performance (depending on how resources are provisioned).


\begin{figure}[t]
    \centering
    \includegraphics[width=0.44\textwidth]{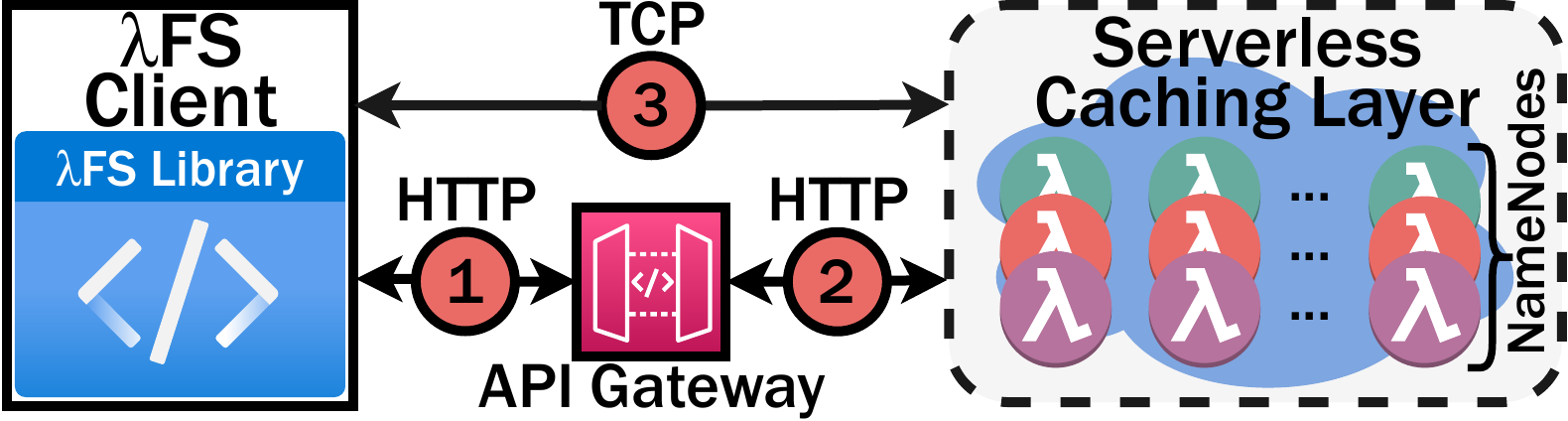}
    \vspace{-4pt}
    \caption{{\proj} supports two different types of metadata RPC requests: HTTP-based and TCP-based RPCs.}
    \label{fig:serverless-hopsfs-requests}
    \vspace{-12pt}
\end{figure}

\subsection{Hybrid Serverless RPC Mechanism}
\label{subsec:design-metadata-requests}

{\proj} utilizes two different RPC pathways in order to provide high system throughput and high elasticity. Specifically, HTTP RPCs directed to the serverless platform's API gateway are used to scale-out the number of serverless function instances as the load increases. At the same time, NameNodes establish direct TCP connections back to clients. Clients can then use this direct connection as a low-latency alternative to HTTP requests. During our experiments, we found that the average end-to-end latency for read operations was 1-2ms for TCP RPCs and 8-20ms for HTTP RPCs. Clients issue TCP RPCs whenever possible due to the significantly lower latency. (TCP RPCs also experience much smaller end-to-end latency variance compared to HTTP RPCs.) 
The lower latencies lead to substantially higher throughput and overall much lower costs due to the reduced overhead for each file system operation.

Figure~\ref{fig:serverless-hopsfs-requests} depicts the two different types of metadata RPCs that {\proj} supports. The first, labeled as step~(1), is a standard HTTP invocation directed to the API gateway of the FaaS framework (e.g., OpenWhisk). At step~(2), the FaaS API gateway will route this request to a serverless function invoker, which will submit the request to 
an existing NameNode, or the invoker will provision a new instance if none exist or all are busy serving other requests. A NameNode that serves an HTTP request will subsequently establish TCP connections back to the clients that issued the HTTP request if no such connection already exists as shown in step~(3).

By default, all clients on the same VM will use the same TCP server (on that VM) to communicate with serverless NameNodes. Users can optionally configure ${\proj}$ to assign at-most $n$ clients to each TCP server. New TCP servers are automatically created for new clients as needed. Clients transparently \diffcomment{include}{inform the NameNodes of} \addcomment{their IP address and the} ports for all TCP servers on their VM \addcomment{within HTTP request payloads}, which enables the NameNodes to proactively connect to \diffcomment{the}{all of these} servers.

\begin{figure}[t]
    \centering
    \includegraphics[width=0.46\textwidth]{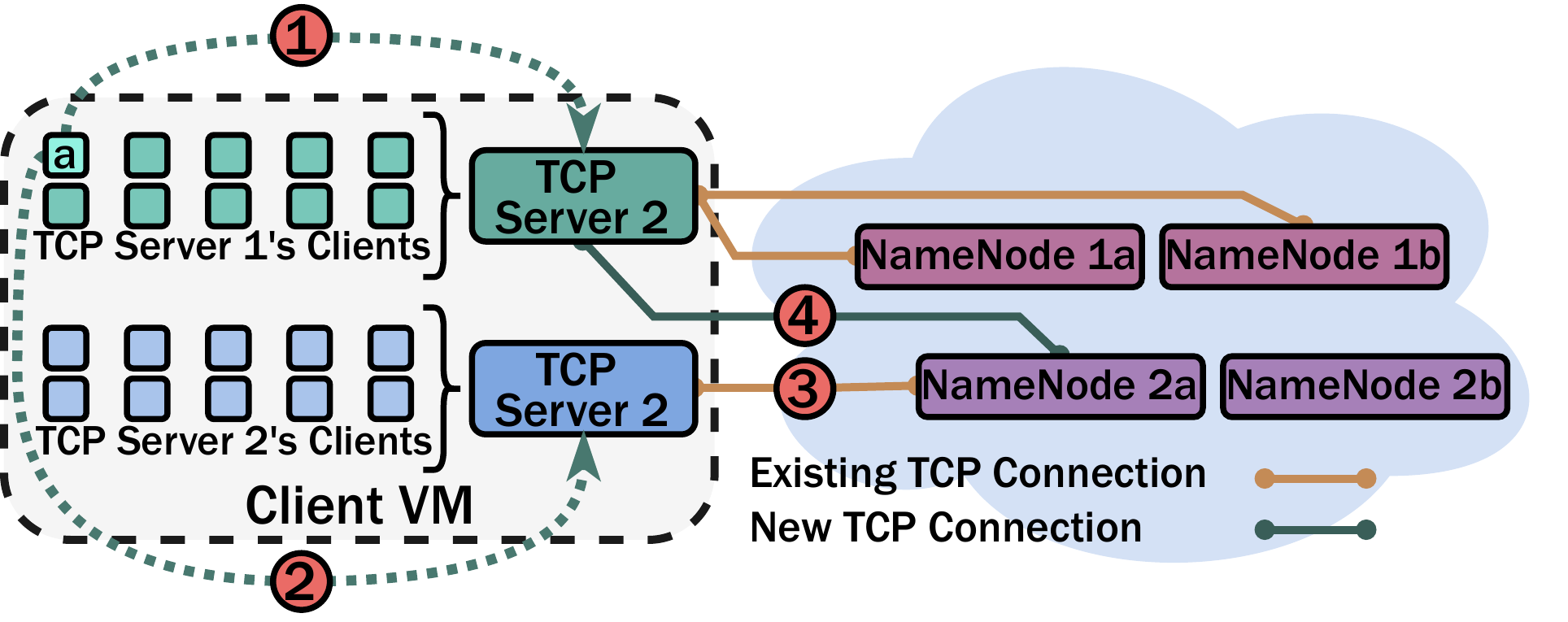}
    \vspace{-6pt}
    \caption{{\proj}' ``connection sharing'' mechanism.}
    \label{fig:connection-sharing}
    \vspace{-10pt}
\end{figure}

Clients also temporarily \textit{share connections} with one another. Consider the process illustrated in Figure~\ref{fig:connection-sharing}. Client $a$ wishes to submit a metadata request to deployment 2 ($D2$). In step 1, $a$ finds that there is no existing connection between its TCP server and an instance of $D2$. Thus, in step 2 client $a$ contacts the other TCP servers on its VM and finds that TCP Server 2 has an existing connection to an instance of $D2$. In step 3, $a$ uses TCP Server 2 to issue its metadata request. After fulfilling the request, NameNode 2$a$ establishes a TCP connection back to client $a$'s assigned TCP server.

When HTTP requests time out, clients could resubmit the requests to the FaaS platform immediately, causing a request storm that could overwhelm the FaaS platform and lead to the over-provisioning of NameNodes. We designed the client library so that clients sleep before resubmitting requests, following an exponential backoff delay pattern with randomized jitter added. 

Similarly, if a TCP connection between a client and a NameNode is dropped, then any incomplete requests are transparently re-submitted by the client. The client will first determine if there are any other active TCP connections to instances from the target deployment. If so, then these connections will be used to re-submit the requests. If not, then the client queries the other TCP servers on its VM, if any, for active connections, and uses any connection it finds to resubmit its request. If no such TCP connections exist, then the client will simply fall-back to HTTP to re-submit the request. Additionally, NameNodes temporarily cache results returned to clients in the event that network delays or other failures prevent the client from receiving the result. When the NameNode receives a re-submitted request, it will attempt to return cached results before re-performing the requested operation.

\subsection{Serverless Metadata Cache}
\label{subsec:design-metadata-cache}


{\proj} NameNodes provide a serverless caching layer for performance.
We partition the file namespace across all the NameNode deployments by consistently hashing on the parent INode ID. Each deployments' NameNode instances are then responsible for caching a partition of the namespace, and clients route metadata RPCs based on this partitioning scheme. 

\begin{figure}[t]
    \centering
    \includegraphics[width=0.46\textwidth]{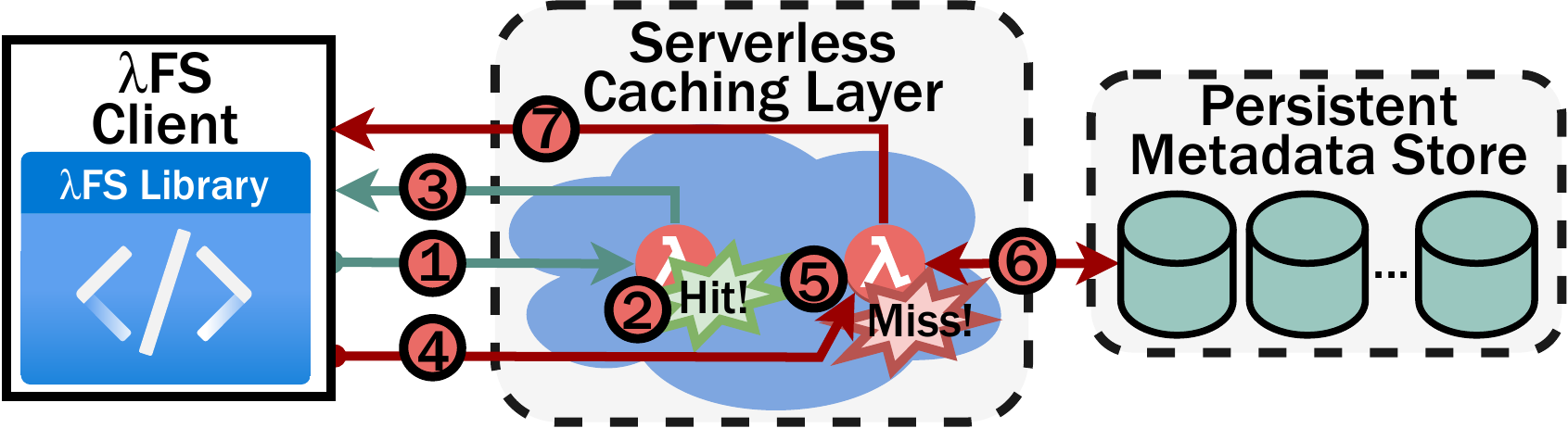}
    \vspace{-4pt}
    \caption{A walkthrough of {\proj}' caching protocol. 
    }
    \label{fig:serverless-hopsfs-cache}
    \vspace{-10pt}
\end{figure}

\begin{figure*}
    \centering
    \includegraphics[width=1\textwidth]{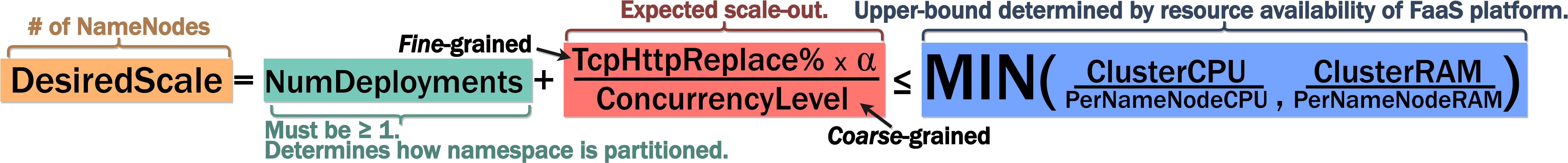}
    \vspace{-10pt}
    \caption{Mathematical model of the agile and lightweight FaaS auto-scaling in {\proj}.}
    \label{fig:scaling-model-equation}
    \vspace{-5pt} 
\end{figure*}

NameNodes cache more than just the metadata associated with the terminal INode in a particular path. Specifically, they cache the metadata for \textit{all} INodes contained within a particular path. Cached metadata is stored in a \textit{trie} data structure maintained in-memory on the NameNode. This caching scheme allows for metadata read operations to avoid going to the persistent metadata store, as NameNodes can serve the read entirely from their local cache, called a cache \emph{hit}.
A cache \textit{miss} occurs when a missing 
metadata must be retrieved from the metadata store. Once retrieved, the NameNode will cache the metadata in its local cache for future read operations. 

Figure~\ref{fig:serverless-hopsfs-cache} provides an illustration of {\proj}' serverless metadata caching. In step~(1), the client issues a metadata request for the file ``{\small\texttt{/nts/notes.txt}}''. 
This results in a cache \textit{hit} on the NameNode in step~(2). As a result, this NameNode can return the metadata directly to the client in step~(3) without having to first retrieve it from the metadata store. Next, the client issues another metadata RPC for the file ``{\small\texttt{/bks/book.pdf}}''. This request is routed to a different NameNode in step~(4) and results in a cache \textit{miss} in step~(5). In step~(6), the consulted NameNode retrieves the metadata from the metadata store. Finally, the metadata is returned to the client in step~(7). 
\subsection{Agile Serverless NameNode Auto-Scaling}
\label{subsec:design-agile-auto-scaling}

\addcomment{A metadata cache reduces per-operation latency and improves system throughput; however, a cache alone is insufficient for supporting large-scale, bursty workloads while maximizing cost-effectiveness and resource efficiency. An MDS equipped with a cache must still be statically provisioned by the user ahead of time. This creates a dilemma for the user, which is to over- or under-provision the resources, trading off performance and cost-efficiency. In order to avoid this trade-off, we implement an agile and lightweight auto-scaling policy for managing coordinated scaling within a serverless framework. We begin by motivating the design of our auto-scaling policy before discussing the general model in detail.}

When clients issue metadata requests, 
they \diffcomment{issue either}{must choose between issuing} a TCP RPC \diffcomment{or}{and} an HTTP RPC (\cref{subsec:design-metadata-requests}). Clients will choose to issue a TCP RPC whenever a 
TCP connection exists to a NameNode in the target deployment. \diffcomment{This is because HTTP RPCs incur significantly higher overhead compared to TCP RPCs; clients only choose HTTP RPCs as a last resort, when no TCP connections exist. However, TCP RPCs are \textit{not} FaaS-aware. Only HTTP RPCs are routed through the FaaS platform and thus enable the platform to detect when additional containers are needed. Therefore, exclusive use of TCP RPCs will ultimately lead to poor scalability and elasticity by preventing auto-scaling, leading to overload and performance degradation.}{However, exclusively relying on TCP RPCs will ultimately lead to poor scalability and elasticity, as TCP RPCs are \textit{not} FaaS-aware. Only HTTP RPCs are routed through the FaaS platform and \diffcomment{thus}{therefore} enable the platform to detect when additional containers are needed. Exclusive use of TCP RPCs tends to result in an inability to auto-scale, and therefore leads to overload and performance degradation.}
This scenario highlights an interesting trade-off between performance (low latency and high throughput) and \diffcomment{elasticity}{scalability/elasticity}. 


To address the 
scenario above, we implement an agile and lightweight auto-scaling policy based on a randomized HTTP-TCP replacement mechanism. 
Each TCP RPC can be probabilistically replaced by an HTTP RPC, with a configurable probability. As the request load increases, the absolute number of HTTP RPCs should increase, enabling the FaaS platform to provision additional serverless containers as needed. In essence, the randomized replacement mechanism allows for a majority of RPCs to be TCP-based while still enabling auto-scaling to occur, leading to better elasticity and scalability while achieving low latency and high throughput.

The auto-scaling policy can be modeled using the equation in
Figure~\ref{fig:scaling-model-equation}, where 
$\alpha$ is a parameter encoding the load level (requests per second and load concurrency), and \textbf{ConcurrencyLevel} is the \textit{function-level concurrency} of each individual NameNode. 
To support asynchronous \textbf{ConcurrencyLevel}, we extended OpenWhisk~\cite{openwhisk} to enable control over how many unique HTTP RPCs a single function instance 
can serve \textit{simultaneously}. This parameter provides \textit{coarse-grained} control over the degree of auto-scaling, as small changes in this value will have a large impact on the number of provisioned NameNodes. The closer that the \textbf{ConcurrencyLevel} is to its minimum value of 1, the greater the degree of auto-scaling. Meanwhile, the HTTP-TCP replacement probability provides \emph{fine-grained} control over auto-scaling. We find that empirically setting the probability of random HTTP-TCP replacement to a value $\leq 1\%$ tends to provide the best performance for the request loads and resource limits we used.
{\proj}' auto-scaling policy reuses the 
FaaS platform's existing auto-scaling facility
while remaining agile and performance-preserving.
We choose not to use sophisticated feedback-based policies, such as Kubernetes' Horizontal Pod Autoscaling algorithm~\cite{k8s_hpa}, as these policies typically require a long feedback-loop delay, which 
cannot be tolerated if sudden load bursts must be dealt with quickly.
\emph{We envision that 
this model is readily applicable to and useful for future performance-sensitive FaaS-based systems. 
\diffcomment{It}{as it} provides an effective methodology that \diffcomment{enables}{makes} traditional FaaS \addcomment{platforms to} embrace high-throughput, low-latency stateful applications.}

\subsection{Coherence Protocol}
\label{subsec:design-consistency-protocol}

Supporting stateful and parallel caching atop serverless NameNodes requires special treatment for concurrent metadata operations, as multiple function instances for the same NameNode deployment may cache replicas of the same metadata. Like HDFS, {\proj}' metadata operations fall into the following two categories: single INode operations that operate on a single file or directory (e.g., 
{\small\texttt{read/create}}
file), and subtree operations that operate on one or more directories spanning many INodes (e.g., recursive {\small\texttt{mv}} and {\small\texttt{delete}}).

Inspired by cache/memory
coherence algorithms~\cite{snoopy_sfcs86, dsm_tocs89},
we designed a modular, serverless memory coherence protocol that guarantees data consistency for DFS metadata. 
The protocol uses a simple {\small\texttt{ACK-INV}} mechanism to ensure that NameNodes have invalidated their caches before any new metadata is persisted to the metadata store. That is, when a NameNode performs a write operation on an INode, it issues an invalidation ({\small\texttt{INV}}) to the 
instances in the
deployment responsible for 
caching each piece of metadata related to the modified INode.
The write operation blocks until all \diffcomment{active}{actively-running} NameNodes have acknowledged ({\small\texttt{ACK'd}}) this \diffcomment{{\small\texttt{INV}}}{invalidation}, at which time the write operation can safely proceed. 
Our coherence protocol utilizes the pluggable ``Coordinator'' service to facilitate communication among the 
NameNodes. The Coordinator is used to keep track of which NameNode instances are actively running in which deployments and to deliver the {\small\texttt{ACKs}} and {\small\texttt{INVs}}. {\proj} builds a subtree coherence protocol atop the simple, single-INode-based protocol  (see Appendix~\ref{subsec:subtree-consist-proto}). 

\begin{algorithm}[tb]
\caption{{\proj} Coherence Protocol}
\label{proto:consistency-protocol}
\SetKwInput{KwInput}{Input}              
\SetKwInput{KwOutput}{Output}            
\DontPrintSemicolon
  \begin{enumerate}[noitemsep,leftmargin=*]
    \item For each $d \in \mathcal{D}, \ N_L$ subscribes to and listens for liveness and {\small\texttt{ACK}} notifications before issuing an {\small\texttt{INV}}, whose payload includes the metadata to be invalidated, to that deployment. All of this is performed using the Coordinator. {\small\texttt{ACK}}s are \textit{not} required from NameNodes that terminate mid-protocol. 


    \item Upon receiving an {\small\texttt{INV}}, NameNodes in each $d \in \mathcal{D}$ first invalidate their caches before responding with an {\small\texttt{ACK}}. 

    

    \item Once $N_L$ has received all required {\small\texttt{ACK}}s, the write operation can safely continue. Metadata changes/updates are persisted to the persistent datastore.
  \end{enumerate}
\end{algorithm}
  

To describe the coherence protocol, we use the following notations. First, recall that there are $n$ deployments across which the NameNodes are partitioned. Let $\mathcal{D}$ denote the set of deployments caching at least one piece of metadata in the target path of a write operation. Next, let $N_L$ denote the ``leader'' NameNode, which is the NameNode performing the write operation. To orchestrate the coherence protocol, $N_L$ actively communicates with 
other NameNodes via the Coordinator.

The coherence protocol is described in Algorithm~\ref{proto:consistency-protocol}. It is conceptually divided into three 
steps. There is also a small amount of clean-up that is performed after the protocol terminates; this step is omitted for simplicity.
The protocol guarantees the serialization of concurrent writes by utilizing exclusive locks in the persistent datastore. First, consider how once the leader NameNode $N_L$ receives all {\small\texttt{ACKs}} from its followers, it is necessarily true that all other NameNodes will have invalidated their caches. Next, $N_L$ will have taken exclusive write-locks on the metadata in the persistent datastore, so it will be impossible for another NameNode to read and cache the metadata before it is updated. This effectively serializes write operations against any other concurrent writes on the same data, thereby guaranteeing strong consistency. 




\subsection{Fault Tolerance}
\label{subsec:design-fault-tolerance}

\addcomment{By default, both single INode and subtree operations do not span multiple NameNodes; however, a multi-node subtree batching mechanism (described in-detail in Appendix~\ref{subsec:subtree-consist-proto}) 
may be enabled to reduce the latency of subtree operations. {\proj} reuses HopsFS' transaction model, and thus both individual request- and NameNode-level failures are handled exactly as HopsFS handles them. Clients transparently resubmit subtree operations to other NameNodes in the event of a crash. In the multi-node case, the failure of any node will be treated as though the entire operation failed, and clients will simply resubmit the operation. Since {\proj}' persistent data store provides ACID transaction semantics, and coupled with {\proj}' consistency protocol, failures cannot leave the namespace in an inconsistent state. Likewise, {\proj}' Coordination service ensures that crashes are detected, enabling the easy removal of locks held by crashed NameNodes.}

\section{Implementation}
\label{sec:implementation}

\noindent\textbf{Implementing \proj.} 
{\proj} is implemented as a fork of HopsFS 3.2.0.3.
Both {\proj}~\cite{lambda-fs-source} and the benchmarking application~\cite{lambda-fs-benchmark-app} are open-sourced. 
{\proj} can be used as a drop-in replacement for HopsFS since {\proj}' client API is a superset of the HopsFS API. {\proj} uses a deployment of Apache OpenWhisk~\cite{openwhisk} as its FaaS platform. 
{\proj} also supports other FaaS platforms including Nuclio~\cite{Nuclio}. Notably, adding support for Nuclio required just 108 additional lines of Java code in {\proj}.
Additionally, {\proj} uses MySQL Cluster NDB 8.0.26 as its persistent metadata store and ZooKeeper~\cite{zookeeper_atc10} as its ``Coordinator'' service.

\begin{table}[t]
\caption{Lines of code required by the different components involved in the development and evaluation of {\proj}.}
\vspace{-6pt}
\label{tab:loc-table}
    \begin{tabular}{lr|lr} 
      
      \hhline{|----|}
      
      \rowcolor[HTML]{d6eaf8} 
      {\bf Component} & {\bf LoC} & {\bf Component} & {\bf LoC} \\
      
      \hhline{|----|}
      
      {Benchmark drivers} & \diffcomment{15,000}{14,000} & {Docker images} & 2,100 \\
      
      \rowcolor[HTML]{EFEFEF} 
      {{\proj}} & \diffcomment{36,685}{35,685} & {Python scripts} & 1,200 \\
      
      \rowcolor[HTML]{EFEFEF} 
      {hammer-bench} & 3,160 &  $\lambda$IndexFS & 4,472 \\
      
      {} & {} & \cellcolor[HTML]{9AFF99}{\textbf{Total:}} & \cellcolor[HTML]{9AFF99}{\textbf{63,624}} 
      
      
    \end{tabular}
\vspace{-10pt}
\end{table}


\begin{figure}[h]
\begin{center}
\vspace{-6pt}
\includegraphics[width=0.48\textwidth]{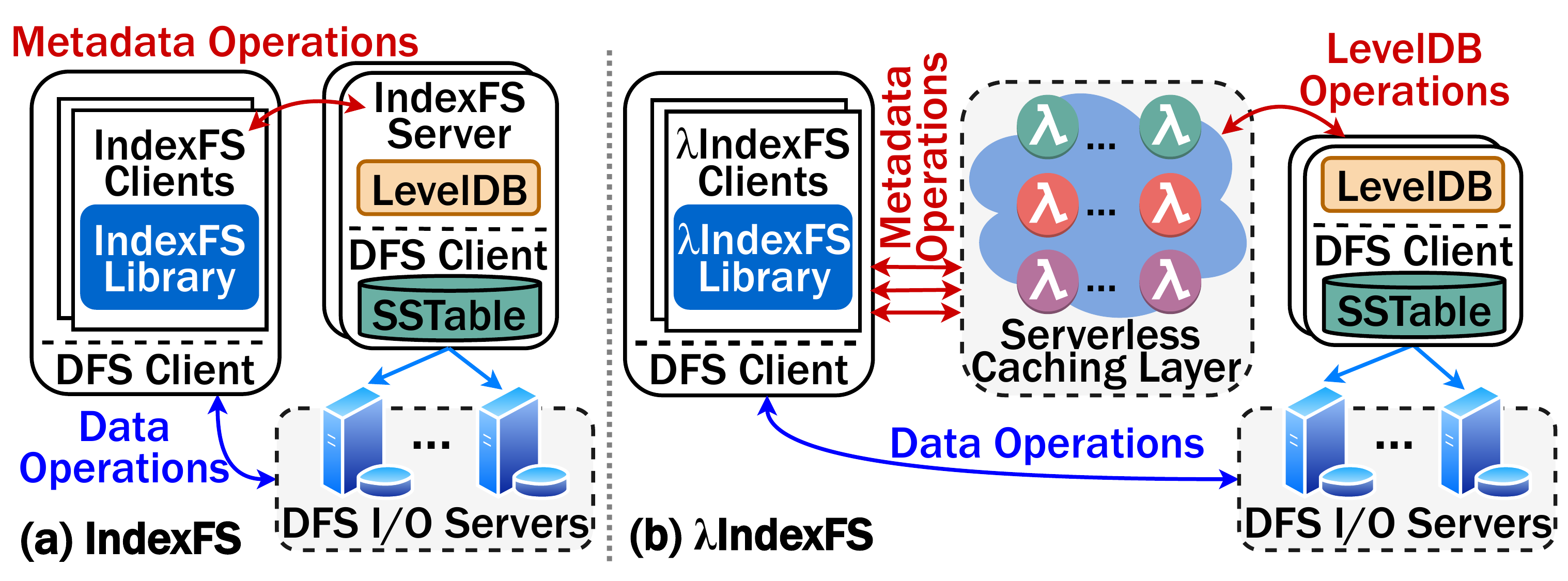}
\vspace{-15pt}
\caption{
Porting {\proj} to IndexFS.
}
\label{fig:IndexFS-vs-lambdaIndexFS}
\end{center}
\vspace{-10pt}
\end{figure}

\noindent
\textbf{Porting {\proj} to IndexFS.}
We have also ported {\proj} to
IndexFS, a scalable middleware MDS~\cite{indexfs_sc14} for DFSes such as BeeGFS~\cite{beegfs}.
Thanks to {\proj}' modular design,  the integration of {\proj} and IndexFS is conceptually similar to that of {\proj} and HopsFS, as  shown in Figure~\ref{fig:IndexFS-vs-lambdaIndexFS}. 
We briefly discuss a few key differences below. 

First, vanilla IndexFS relies on LevelDB to pack metadata 
into  SSTables~\cite{leveldb}. We decouple in-memory metadata handling from backend LevelDB by packaging
the logic into serverless functions, and only using LevelDB as the persistent metadata store.
Second, IndexFS leverages a sophisticated 
metadata partitioning algorithm adapted from 
GIGA+~\cite{giga+_fast11}. 
After discussions with the IndexFS authors, we developed an alternative partitioning scheme that is easier to integrate with {\proj}. This scheme uses hashing to partition directories
{across LevelDB SSTables} by directory names.  
Third, to make C++ based IndexFS compatible with {\proj}' Java-based serverless functions, we addressed multiple engineering challenges involving cross-language data types 
and library compatibility (e.g., Java's KryoNet~\cite{kryonet} is not available for C++).
Overall, we find that porting {\proj} is managable as {\proj} is designed to be modular and generalizable. 
For simplicity, we refer to our {\proj}-ported IndexFS as {\lindexfs}.

\smallskip
\noindent\textbf{Porting {\proj} to Commercial FaaS Platforms.}
It is straightforward to port {\proj} to commercial FaaS platforms such as AWS Lambda.
{\proj}' core techniques are not dependent on any particular FaaS platform. This includes {\proj}' RPC mechanism, as other frameworks have successfully used TCP-RPC-like mechanisms on commercial FaaS platforms in the past~\cite{excamera_nsdi17}. {\proj} could in theory be deployed on any FaaS platforms that support custom-container-based function deployment~\cite{fc_custom_container,google_cloud_run,aws_lambda_container}.
One challenge is how to minimize the performance impact of warm function reclamation~\cite{peeking_atc18}, which we leave as our future work. 



\smallskip
\noindent\textbf{Summary of Implementation Efforts.}
We have implemented {\proj} and the software used for its evaluation in roughly \diffcomment{63,624}{61,624} lines of Java/C++ code (see Table~\ref{tab:loc-table}), completed over the course of more than two person-years. The benchmarking software constitutes \diffcomment{18,160~LoC}{17,160~LoC}, while {\proj} and {\lindexfs} together are composed of approximately \diffcomment{41,157~LoC}{40,157~LoC}.

\section{Evaluation}
\label{sec:eval}

\subsection{Experimental Setup \& Methodology}
\label{subsec:eval_methodology}

\addcomment{In order to elucidate the effectiveness of {\proj}' various techniques and optimizations, we evaluated {\proj} against a number of other file systems. Specifically, we compared {\proj} against three state-of-the-art distributed file systems: HopsFS, IndexFS~\cite{indexfs_sc14}, and CephFS~\cite{ceph_mds_sc04}. We also performed experiments that evaluated {\proj}' performance against that of a modified HopsFS, denoted ``HopsFS+Cache'', whose NameNodes had been augmented with an in-memory metadata cache similar to that of {\proj}. HopsFS+Cache serves as a serverful, cache-based DFS baseline. Finally, we compared {\proj} with {\infcache}~\cite{infinicache_fast20}, an in-memory object cache implemented atop FaaS, in order to better understand the efficacy of {\proj}' FaaS caching layer. 
{\infcache} uses a static, fixed-size deployment of cloud functions to serve I/O operations via short TCP connections that require invoking functions for every operation. {\infcache} thus serves as an approximation of {\proj} with no auto-scaling or long-lived TCP-RPC request mechanism.
All experiments were performed on Amazon Web Services (AWS), and results were verified to be consistent with results obtained on Google Cloud Platform (GCP).}

\addcomment{The experiments used deployments of the OpenWhisk serverless platform and MySQL Cluster NDB. Like {\proj}, HopsFS uses MySQL Cluster NDB as its persistent metadata store. OpenWhisk was deployed on AWS Elastic Kubernetes Service (EKS). All other VMs were deployed on AWS EC2. All AWS VMs used the \texttt{\small{r5.4xlarge}} instance type (16 vCPU and 128GB RAM). MySQL Cluster NDB 8.0.26 was deployed on GCE and EC2 with a single master node and four data nodes. We configured each NDB storage node according to the sample configuration provided by HopsFS. Unless otherwise specified, 
each {\proj} NameNode was configured with 6.25 vCPU and 30GB RAM. HopsFS' NameNodes were configured with 16 vCPU, 64GB RAM, and 200 RPC handlers.} For clarity, we defer the description of the \diffcomment{setup for the portability experiment}{portability experiment setup} with IndexFS to \cref{sec:indexfs-results} as IndexFS' architecture is different from HopsFS.

To ensure a fair comparison between {\proj} and HopsFS, we allocated an equal amount of vCPUs and RAM to each framework's NameNode cluster (unless otherwise specified). \addcomment{However, imposing a fixed, total vCPU limitation on {\proj}' NameNodes implicitly restricted the maximum performance of {\proj} \diffcomment{compared to what {\proj} could have achieved with}{that {\proj} could achieve compared to {\proj}' performance with} the nearly unbounded resources typically provided by FaaS platforms.}
Because of this \addcomment{self-imposed} bound \addcomment{on vCPUs},
unrestricted {\proj} scale-outs would have over-used resources, leading to thrashing and severe performance degradation. To \diffcomment{prevent}{address} this, {\proj}' scaling behavior was ``toned down'', and consequently {\proj} never \diffcomment{actively provisioned}{used} more than \diffcomment{92.77\%}{79\%} of the available vCPUs during these experiments. (We \diffcomment{describe}{introduce} {\proj}' anti-thrashing technique in Appendix~\ref{appendix:anti-thrashing}.) 
The resource scaling tests presented in \cref{subsubsec:strong-scaling} illustrate how {\proj}' performance improves as more resources are allocated to {\proj}, thereby providing insight into how {\proj} would perform with nearly unbounded resources. 

Our evaluation aims to answer the following questions:
\begin{itemize}[noitemsep,leftmargin=*]

    \item \diffcomment{How does {\proj} perform under industrial workloads (\cref{subsec:real-world-workload})?}
        
    \item \diffcomment{To what extent does {\proj}' elasticity improve performance, scalability, resource- efficiency, and cost-efficiency  (\cref{subsubsec:autoscaling}, \cref{subsec:as}, \cref{subsubsec:real-world-cost}, \cref{subsubsec:cost-efficiency-microbenchmarks})?} 
    
    \item How does {\proj} scale for individual DFS operations \addcomment{compared to other large-scale DFSes} (\cref{subsec:eval-scaling})? 

    \item \addcomment{Is {\proj} resilient to serverless NameNode failures (\cref{subsec:ft})?} 
    
    
    \item Can \proj benefit other DFSes besides HopsFS (\cref{sec:indexfs-results})? 
    
    
\end{itemize} 

\subsection{Industrial Workload}
\label{subsec:real-world-workload}

In this section, we present and discuss the results of executing a real-world, industrial workload on both {\proj} and HopsFS. The workload is based on the one used in HopsFS' evaluation, which was generated using statistics from traces of Spotify's 1600-node HDFS cluster. The frequencies of the file system operations are shown in Table~\ref{tab:spotify-operation-percentages}. 

\begin{table}[t]
\caption{Relative frequency of the file system operations used in the Spotify workload experiment.}
\vspace{-6pt}
\begin{tabular}{@{}lr|lr@{}} 
\rowcolor[HTML]{d6eaf8} 
\textbf{Operation} & \textbf{Percentage} & \textbf{Operation}                & \textbf{Percentage}             \\
{\small\texttt{create}} file & 2.7\%  & {\small\texttt{read}} file     & 69.22\% \\
\rowcolor[HTML]{EFEFEF} 
{\small\texttt{mkdirs}}      & 0.02\% & {\small\texttt{stat}} file/dir & 17\%    \\
{\small\texttt{delete}} file/dir  & 0.75\% & {\small\texttt{ls}} file/dir & 9.01\%  \\
\rowcolor[HTML]{EFEFEF} 
{\small\texttt{mv}} file/dir             & 1.3\%               & \cellcolor[HTML]{9AFF99}\textbf{Total Read Ops} & \cellcolor[HTML]{9AFF99}\textbf{95.23\%}
\end{tabular}
\label{tab:spotify-operation-percentages}
\vspace{-10pt}
\end{table}


\begin{figure*}[t]
\begin{center}
\subfigure[Base: 25,000 ops/sec.] {
\includegraphics[width=0.333\textwidth] {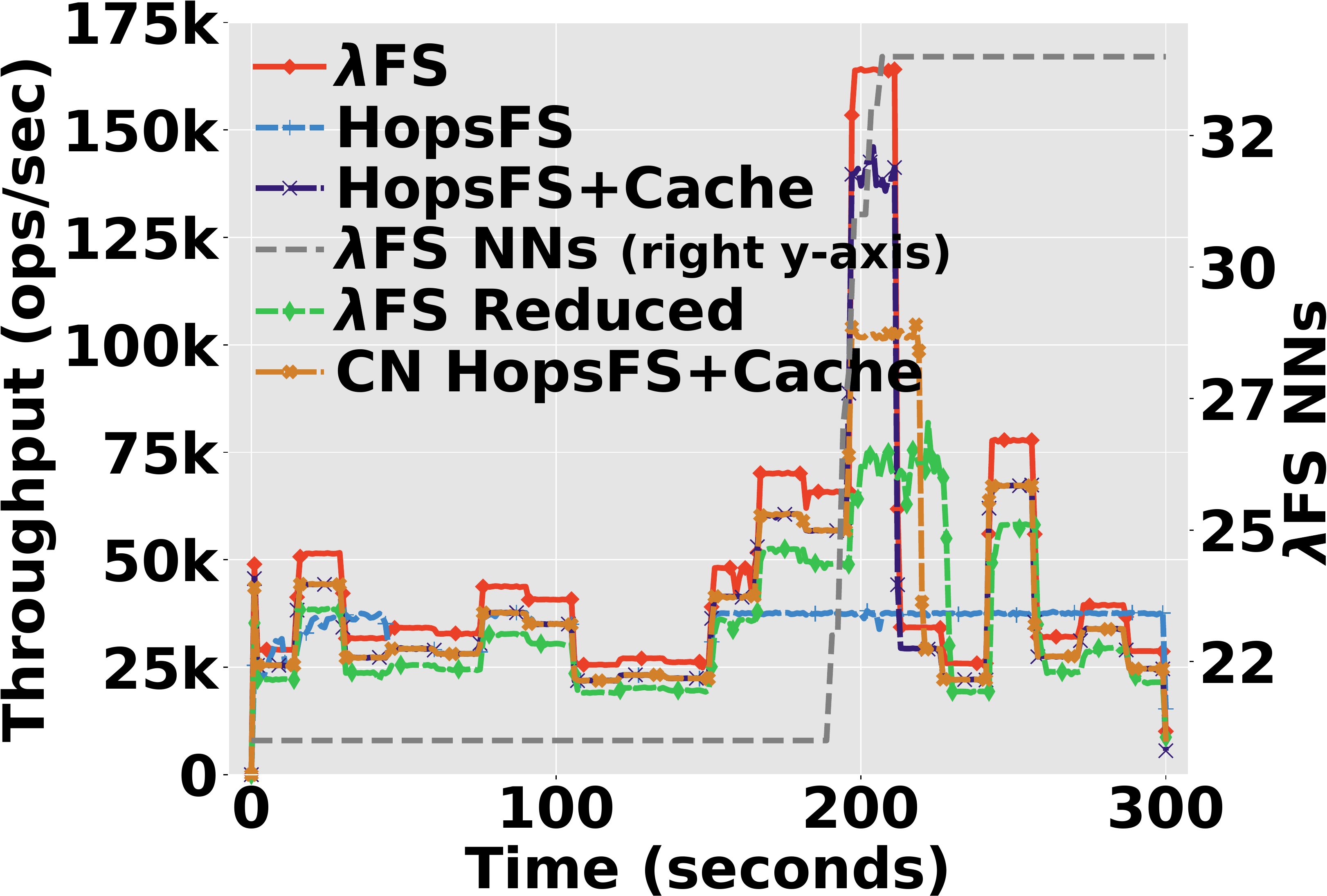}
\label{fig:spotify-25k-throughput-new}
}
\hspace{-2pt}
\subfigure[Base: 50,000 ops/sec] {
\includegraphics[width=0.30\textwidth]{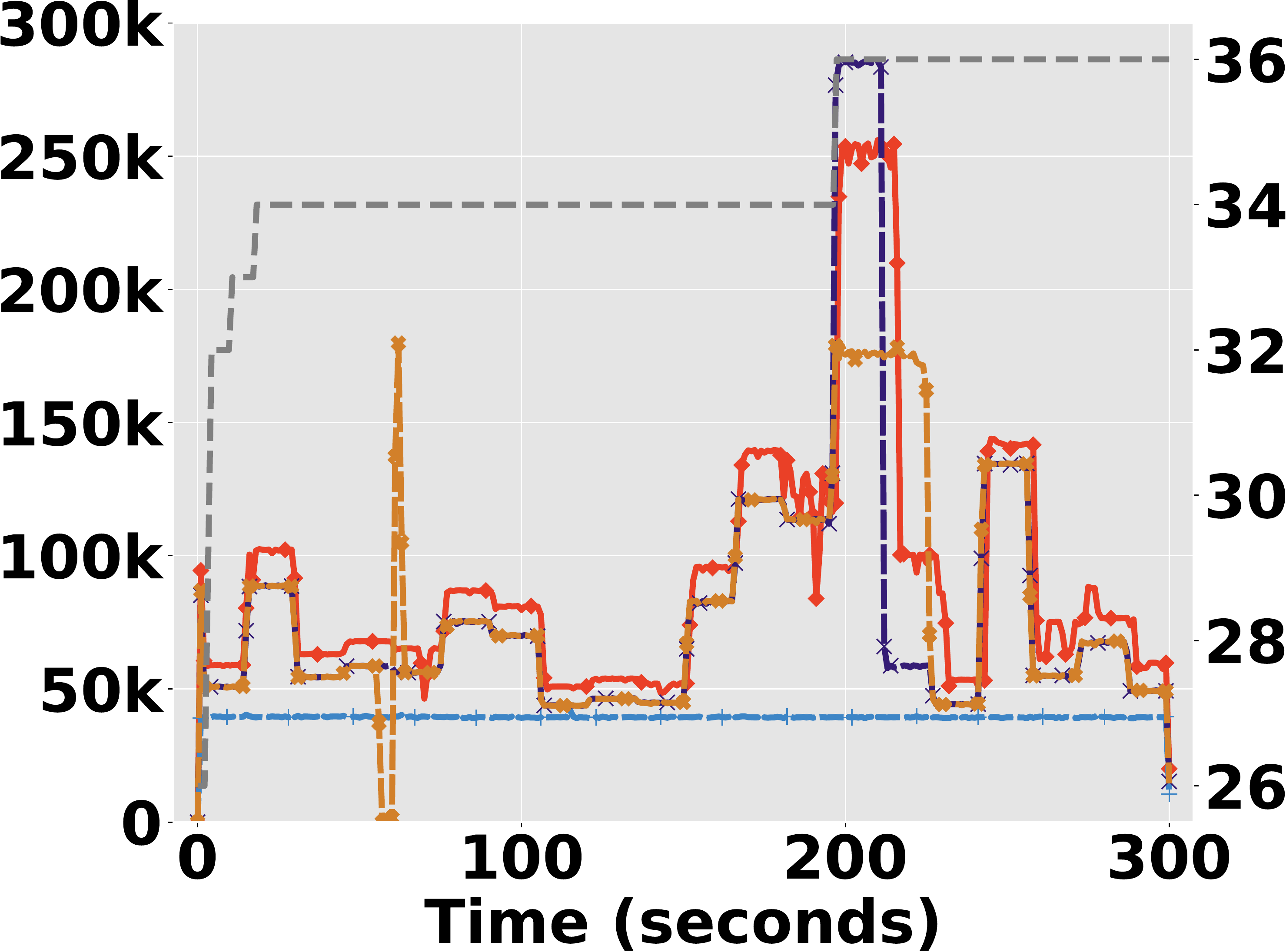}
\label{fig:spotify-50k-throughput-new}
}
\hspace{-2pt}
\subfigure[Performance-per-cost comparison.] {
\includegraphics[width=0.305\textwidth] {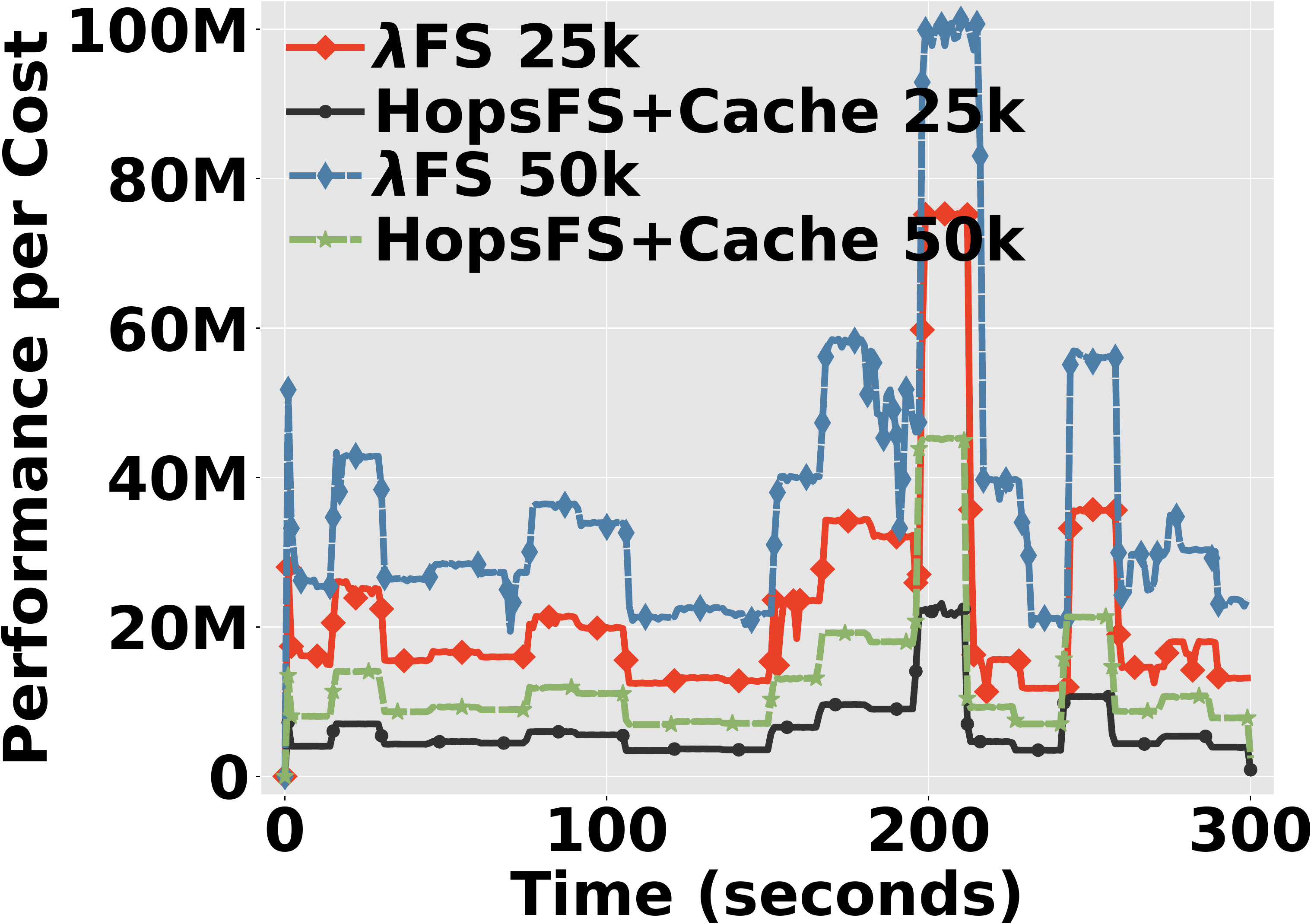}
\label{fig:spotify-perf-per-cost-new}
}
\vspace{-12pt}
\caption{Throughput and performance-per-cost comparison between \diffcomment{the various systems}{{\proj} and HopsFS} during the Spotify workload. 
The number of active {\proj} NameNodes (``NNs'') is shown on the secondary $y$-axis in both Figure~\ref{fig:spotify-25k-throughput-new} and \ref{fig:spotify-50k-throughput-new}. 
}
\label{fig:spotify-throughput-and-perf}
\vspace{-5pt}
\end{center}
\end{figure*}

\begin{figure}[t]
    \centering
    \includegraphics[width=0.45\textwidth]{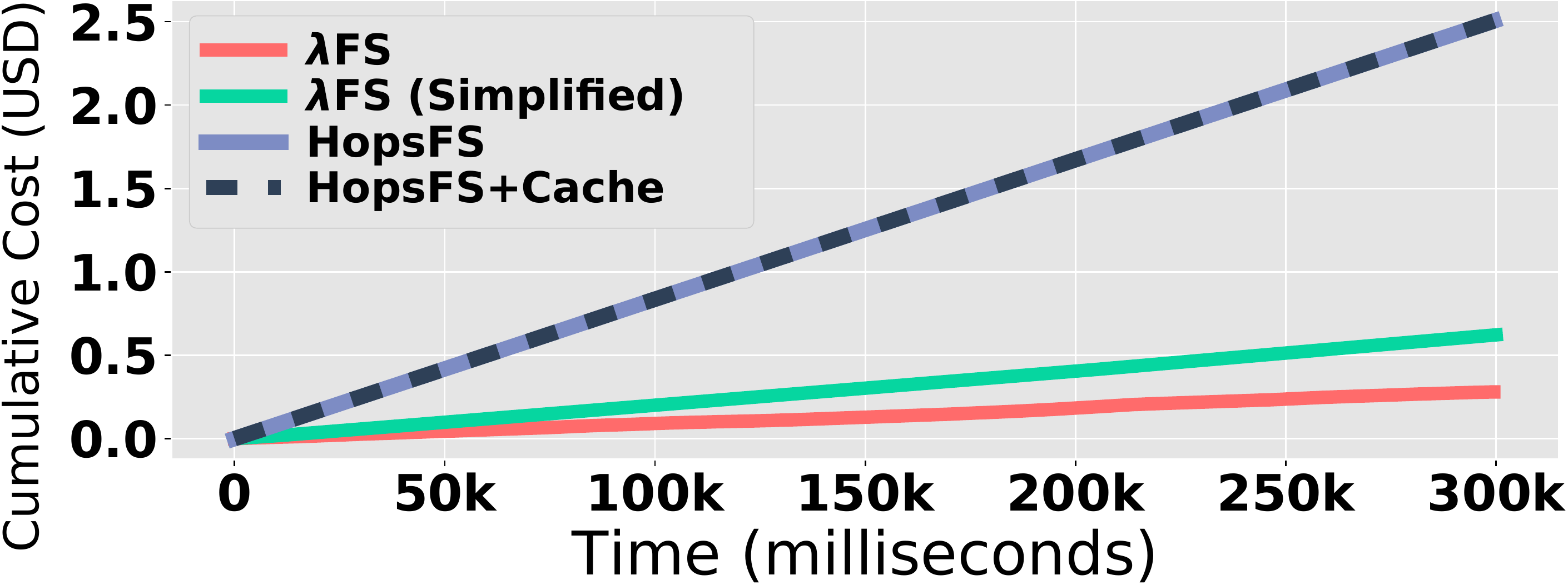}
    \vspace{-8pt}
    \caption{Cumulative cost of the 25k ops/sec Spotify workload. HopsFS' cost was \$2.50. {\proj}' cost was \$0.35 using AWS Lambda's prices, which are \$0.0000166667 per GB-second, charged at 1ms granularity, and \$0.20 per 1M requests~\cite{lambda_pricing}. \addcomment{Under the ``simplified'' cost model, {\proj} NameNodes incur cost while they're provisioned, similar to VMs, which overcharges compared to AWS Lambda's pay-per-use pricing model.}
    }
    \vspace{-10pt}
    \label{fig:spotify-25k-cost-cumulative}
\end{figure}

\begin{figure*}[t]
\begin{center}
\includegraphics[width=1.0\textwidth]{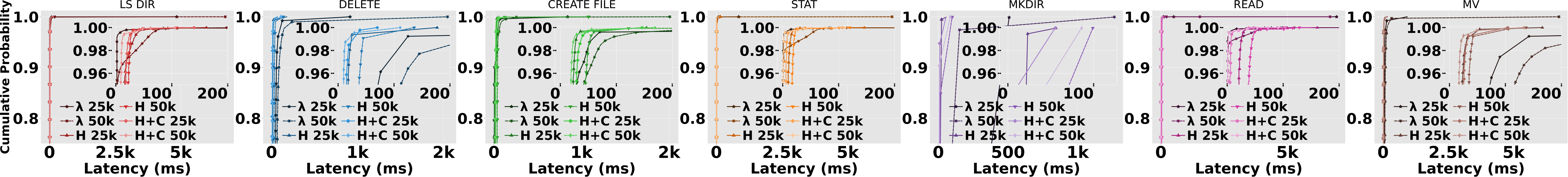}
\vspace{-15pt}
\caption{Latency CDFs 
\diffcomment{of {\proj} (``$\lambda$''), HopsFS (``H''), and HopsFS+Cache (``H+C'')}{(``$\lambda$'') and HopsFS (``H'')} for \diffcomment{both}{the two} versions of the Spotify workload.}
\label{fig:spotify-latency-cdf}
\vspace{-10pt}
\end{center}
\end{figure*}

\begin{figure*}[t]
\begin{center}
\includegraphics[width=1.0\textwidth]{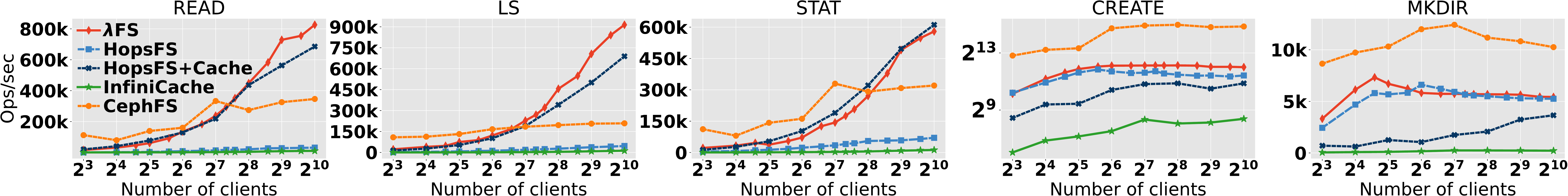}
\vspace{-15pt}
\caption{\textit{Client-driven} scaling comparison between \diffcomment{the various systems}{{\proj} and HopsFS}. The total amount of vCPUs allocated to {\proj} and HopsFS was held constant at 512 vCPUs, and each client performed \diffcomment{3,072}{2,048} operations. The number of clients ranged from 8 to 1,024. \addcomment{Note that the $y$-axis for the {\texttt{\small{create}}} operation is log-scale.}}
\label{fig:weak_scale_throughput}
\vspace{-5pt}
\end{center}
\end{figure*}

\vspace{-4pt}
\subsubsection{Experimental Setup}
We implemented a DFS benchmark that can generate bursty file system loads by modifying the \texttt{\small{hammer-bench}} utility used to conduct HopsFS' evaluation~\cite{hammer-bench, hopsfs_fast17}.
Our benchmark randomly varies system throughput over the course of the workload's execution in order to accurately simulate a real-world DFS workload~\cite{igen_cloud16}. Specifically, \diffcomment{the workload is executed for 5~minutes.}{our benchmark executes the workload for 5~minutes.} Every 15~seconds, the benchmark generates a random throughput value $\Delta$ from a Pareto distribution with a shape parameter $\alpha=2$. (Please refer to~\cite{igen_cloud16} for a discussion on why the Pareto distribution is useful in this scenario.) 
Each client VM will attempt to sustain $\delta = \frac{\Delta}{n}$ ops/sec, where $n$ is the total number of client VMs. If less than $\delta$ operations are completed \diffcomment{in}{during} a given second, then the remaining operations roll over to the next second. 

In order to demonstrate {\proj}' ability to elastically scale in response to bursts of metadata requests, the benchmark randomly generates throughput spikes up to $7\times$ greater than the \emph{base} throughput. We ran 2 different versions of the workload: one in which the Pareto distribution's scale parameter $x_t=25,000$ and the other in which $x_t=50,000$. The value of $x_t$ determines the \diffcomment{workload's base throughput}{base throughput of the workload}. 
Both workloads were executed by 1,024 clients across 8 VMs. We allocated 512 vCPUs to HopsFS in order to maximize its performance during these tests. While allocating less vCPUs would've been cheaper, HopsFS' performance would've suffered, as shown by the resource scaling tests. Each {\proj} NameNode was allocated 5 vCPUs, and in the 25,000 ops/sec workload, {\proj}' NameNode cluster was collectively allocated just 50\% of the total vCPU allocated to HopsFS' NameNode cluster in order to better illustrate {\proj}' resource and cost efficiency.

\vspace{-4pt}
\subsubsection{Throughput \& Latency}

\addcomment{Figure~\ref{fig:spotify-25k-throughput-new}} shows a throughput comparison between {\proj}, HopsFS, and HopsFS+Cache during an execution of the Spotify workload with a base throughput of 25,000~ops/sec. Note that each of {\proj}' NameNodes was configured with 6GB of RAM for the 25,000 ops/sec workload. {\proj} achieved an average throughput of 45,690.34~ops/sec and an average latency of 1.02~ms
during the execution of this workload. HopsFS achieved an average throughput of 38,134.35~ops/sec and an average latency of 10.58~ms. 
\addcomment{HopsFS+Cache achieved an average throughput of 45,945.1032~ops/sec and an average latency of 3.348~ms.}
Summarily, {\proj} achieved 90.40\% (10.41$\times$) lower latency and 16.53\% (1.19$\times$) higher throughput on average than HopsFS, while using 39.45\% less resources. \addcomment{Compared to HopsFS+Cache, {\proj} achieved equivalent average throughput and 69.53\% ($3.28\times$) lower latency on average.}
{\proj} was successfully completed the entire workload, including the entire 15-second 163,996~ops/sec burst generated \diffcomment{at}{around} time $t=200$. Meanwhile, HopsFS' clients struggled to sustain loads above 38,000~ops/sec. When the burst occurred, HopsFS had already ``fallen behind'' and was struggling to execute operations 
generated nearly a minute prior. So, {\proj}' peak sustained, throughput was 4.3$\times$ higher than that of HopsFS.

Figure~\ref{fig:spotify-50k-throughput-new} shows a throughput comparison 
during an execution of the Spotify workload with a base throughput of 50,000~ops/sec. For this test, {\proj}' FaaS platform was allocated 512 vCPU but used at most 180/512 (35.15\%) of the available vCPUs. {\proj} achieved an average throughput of 90,875.60~ops/sec and an average latency of 4.31~ms during 
this workload. HopsFS achieved an average throughput of 44,956.28~ops/sec and an average latency of 22.40~ms. 

HopsFS was unable to achieve the base throughput of 50,000~ops/sec, so it spent the duration of the workload attempting to ``catch up''. Meanwhile, {\proj} sustained approximately 250,000 ops/sec during the burst at around $t=200$. To this end, {\proj}' peak, sustained throughput was 456.09\% (5.56$\times$) higher than that of HopsFS.

{\proj}' average throughput was 102.14\% (2.02$\times$) greater than HopsFS'; similarly, {\proj}' average latency was 80.76\% (5.19$\times$) lower than HopsFS'. For ``read'' operations, {\proj} achieved an average latency anywhere from $6.93\times$---$20.13\times$ lower than that of HopsFS (see Figure~\ref{fig:spotify-latency-cdf}). However, {\proj} was unable to complete ``write'' operations as quickly as HopsFS because of the added overhead required by {\proj}' coherence protocol. Summarily, HopsFS achieved 1.5$\times$---$5.55\times$ shorter ``write'' latencies compared to {\proj}. 

\addcomment{{\infcache} failed to complete either of the two Spotify workloads. The FaaS platform became overwhelmed by the volume of HTTP requests:
the high-latency HTTP requests and static, fixed-size deployment were insufficient for both the base throughput and bursts of work during the workloads.}

\addcomment{Because FaaS assumes a near-unbounded amount of resources, fixing the amount of vCPU allocated to the platform results in poor performance and scalability. To perform a fair comparison and highlight the cost-saving benefits of FaaS, we also compared {\proj} against a ``cost-normalized'' configuration of HopsFS+Cache, referred to as ``CN HopsFS+Cache''. 
Specifically, we configured CN HopsFS+Cache with $72$ and $144$ vCPU for the 25,000 and 50,000 ops/sec Spotify workloads, respectively. In doing so, CN HopsFS+Cache incurred the same monetary cost as {\proj}. Considering first the 25,000 ops/sec workload as shown in Figure~\ref{fig:spotify-25k-throughput-new}, CN HopsFS+Cache achieved lower throughput than {\proj}, failing to sustain the burst of requests around the 200th second of the workload. This phenomenon occurs again during the 50,000 ops/sec workload as shown in Figure~\ref{fig:spotify-50k-throughput-new}.} 

\vspace{-2pt}
\subsubsection{In-Memory Metadata Cache}
To measure the performance impact of {\proj}' metadata caching layer, we executed \diffcomment{another}{a second} instance of the $x_t=25,000$ workload in which we decreased the capacity of the serverless NameNode cache to less than half the working set size (WSS) of the workload. 
As shown in Figure~\ref{fig:spotify-25k-throughput-new}, 
``reduced-cache {\proj}'' achieved better performance than HopsFS, sustaining between 70,000---80,000~ops/sec during the largest burst. Despite failing to sustain 163,996~ops/sec, \addcomment{``reduced-cache {\proj}''} \diffcomment{quickly caught up and completed}{was able to quickly catch up and complete} the remainder of the workload.  

\vspace{-2pt}
\subsubsection{Elastic Auto-Scaling}
\label{subsubsec:autoscaling}

The results of the Spotify workload demonstrate 
{\proj}' ability to handle large bursts of work. 
Figures~\ref{fig:spotify-25k-throughput-new} and~\ref{fig:spotify-50k-throughput-new} show that {\proj} 
provisioned additional NameNodes to satisfy the influx of requests 
as soon as the workload started. {\proj} quickly scaled-out again \diffcomment{near}{around} the 200-second mark, \diffcomment{which}{as this} is when the 7$\times$ request burst occurred, demonstrating the effectiveness of {\proj}' auto-scaling policy. 
With unbounded resources, {\proj} could rapidly scale-out to much higher load spikes. \addcomment{This is supported by the trend shown in the resource scaling experiments (Figure~\ref{fig:strong_scale_throughput}).}

\vspace{-2pt}
\subsubsection{Monetary Cost}
\label{subsubsec:real-world-cost}

Figure~\ref{fig:spotify-25k-cost-cumulative} shows the cumulative cost \diffcomment{for}{during} the 25,000~ops/sec Spotify workload for {\proj}, HopsFS, \addcomment{and HopsFS+Cache}. 
For {\proj}, the cost was computed as follows: for every 1~ms interval of the workload, we billed each NameNode actively serving an HTTP or TCP request using AWS Lambda's prices (as described in Figure~\ref{fig:spotify-25k-cost-cumulative}). If no requests were actively being served by a particular NameNode, then that NameNode incurred no cost. For HopsFS \addcomment{and HopsFS+Cache}, we billed the cost of the entire 512-vCPU cluster for each 1~ms interval of the workload.
By the end of the workload, the cumulative cost of HopsFS and HopsFS+Cache was \$2.50 while {\proj}' cumulative cost was just \$0.35. By taking advantage of FaaS' pay-per-use pricing model \addcomment{and our agile auto-scaling policy}, {\proj} reduced the cost of executing the workload by 85.99\% (7.14$\times$) compared to HopsFS \addcomment{and HopsFS+Cache} while achieving better performance \addcomment{with fewer resources}.

\addcomment{We also computed the cost of {\proj} using a \emph{simplified} pricing model, which is shown in Figure~\ref{fig:spotify-25k-cost-cumulative} as ``{\proj} (Simplified).'' Under this model, active NameNode instances incurred cost as long as they are provisioned~\cite{provisioned_lambda}, which doubled the cost of {\proj} compared to the pay-per-use FaaS pricing model.
This illustrates how {\proj} takes advantage of FaaS' pay-per-use pricing model to greatly reduce tenant-side costs.} 

\addcomment{While the use of FaaS can yield improved elasticity, scalability, and performance, other primary benefits of FaaS are reduced tenant-side cost and increased cost-effectiveness. In particular, {\proj}' cost-effectiveness arises from its ability to achieve superior or equivalent performance while using a smaller amount of resources. By saturating a large number of relatively small, individual serverless NameNodes, {\proj} exhibits high resource utilization and resource efficiency with respect to the resources provisioned to it by the FaaS provider. Likewise, by leveraging the pay-per-use property of FaaS, {\proj} is ultimately able to reduce workload costs while delivering equivalent or better performance.}

\addcomment{To quantify this notion of cost-efficiency, we define a new metric \textit{performance-per-cost}, given as one of $\frac{\text{instantaneous throughput}}{\text{instantaneous cost}}$ or $\frac{\text{average throughput}}{\text{total cost}}$. The units are $\frac{operations/second}{\$} \big(= \frac{operations}{second \times \$}\big)$, or operations-per-second-per-dollar. 
This metric provides a measurement of cost-efficiency --- a higher value indicates that the associated framework is able to achieve a higher performance-cost ratio, which is desirable. Increasing the value of this metric can be done using some combination of increasing throughput and decreasing cost.}

\addcomment{Figure~\ref{fig:spotify-perf-per-cost-new} shows the performance-per-cost for each second of the real-world Spotify workload for {\proj} and HopsFS+Cache. The cost for HopsFS+Cache is computed as the cost of running the 32 NameNode VMs for one second, whereas the cost for {\proj} is calculated using the pay-per-use pricing model of FaaS. Specifically, the resources allocated to an active NameNode are only billed if that NameNode served a request within that second. {\proj} achieved significantly higher performance-per-cost compared to HopsFS+Cache. This is because {\proj} experienced equal or greater throughput compared to HopsFS+Cache for the entirety of the workload while using significantly fewer resources (at-most 165 or 180 vCPU for {\proj}, depending on the workload, compared to the 512 vCPU used by HopsFS+Cache during both workloads).}
\subsection{Scalability}
\label{subsec:eval-scaling}

Next, we evaluate the scalability of {\proj}, HopsFS, \addcomment{HopsFS+Cache, {\infcache}, and CephFS} using two micro-benchmarks covering key DFS operations including \texttt{\small{read}}, \texttt{\small{stat file}}, \texttt{\small{ls}}, \texttt{\small{mkdir}}, and \texttt{\small{create file}}. \addcomment{All operations target random files and directories across an existing directory tree.}
The first microbenchmark tests {\proj}' \textit{client-driven scaling}: the ability of {\proj} to automatically scale-out as the number of clients increases, given a fixed resource cap.
The second microbenchmark, which we call \textit{resource scaling}, tests horizontal scalability (i.e., performance scaling with more deployments) and intra-deployment, vertical auto-scaling. 
The results of these tests illustrate {\proj}' ability to transparently adapt to increases in both request load and available resources in order to maximize performance. 
We allocated a maximum of 512 vCPUs to \diffcomment{all}{both} systems during these tests, and for {\proj} we provisioned at-most \diffcomment{76}{81} NameNodes, each with \diffcomment{6.25}{5} vCPUs, meaning {\proj} used at-most \diffcomment{$76 \times 6.25$ $=475/512$ (92.77\%)}{$81 \times 5$ $=405/512$ (79\%)} of its allocated vCPUs during these tests. 

\begin{figure*}[t]
\begin{center}
\includegraphics[width=1.0\textwidth]{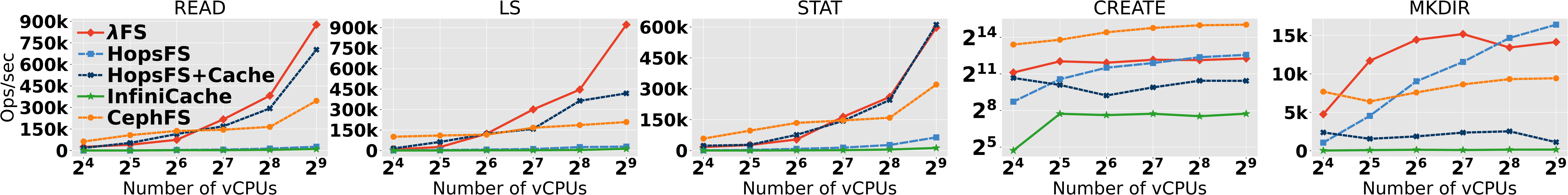}
\vspace{-15pt}
\caption{\textit{Resource scaling} comparison between \diffcomment{the various systems}{{\proj} and HopsFS}. The amount of vCPUs allocated to \diffcomment{the systems}{{\proj} and HopsFS} ranged from 16 \diffcomment{to}{and} 512. \addcomment{For each problem size}, all systems used the same number of clients, each of which performed \diffcomment{3,072}{2,048} operations.}
\vspace{-5pt}
\label{fig:strong_scale_throughput}
\end{center}
\end{figure*}

\vspace{-2pt}
\subsubsection{Client-Driven Scaling}
\label{subsubsec:weak-scaling}

In this test, the amount of vCPUs allocated to both systems was fixed at 512 vCPUs to maximize performance, and each client executed \diffcomment{3,072}{2,048}  operations. The total number of clients configured for each framework was varied between 8 and 1,024 in order to provide a wide range of scales to evaluate the frameworks. The results of this experiment are shown in Figure~\ref{fig:weak_scale_throughput}.  
%

{\proj} \addcomment{ultimately} achieved higher throughput for all read operations (i.e., \texttt{\small{read}}, \texttt{\small{stat}}, and \texttt{\small{ls}}) for all problem sizes:   
{\proj} averaged 
\diffcomment{$28.91\times$}{$21.67\times$},
\diffcomment{$8.22\times$}{$6.44\times$}, and 
\diffcomment{$20.53\times$}{$3.62\times$} higher throughput than HopsFS for \texttt{\small{read}}, \texttt{\small{stat}}, and \texttt{\small{ls}}, respectively. \addcomment{CephFS outperforms the other file systems for \texttt{\small{read}}, \texttt{\small{ls}}, and \texttt{\small{stat}} for the first 4-5 problem sizes but fails to scale well beyond this point, only outperforming HopsFS and {\infcache}. {\proj} outperforms HopsFS+Cache for \texttt{\small{read}} and \texttt{\small{ls}} and achieves comparable performance for \texttt{\small{stat}}.} 

There are several reasons for the throughput differences.
First, {\proj}' elastic caching layer efficiently serves metadata to clients from the memory of serverless functions rather than the persistent metadata store that is 2 network hops away. This also decreases the likelihood of the persistent metadata store becoming a bottleneck,
as it is with HopsFS. 
Second, {\proj} \diffcomment{elastically scales-out}{is able to elastically scale-out} its \diffcomment{NameNodes}{function deployments} \addcomment{in accordance with its agile auto-scaling policy in order} to satisfy the increasing request load, whereas HopsFS is limited by its fixed-scale deployment. 
Under {\proj}, there were 20 active NameNode instances for the 8 client case \addcomment{during the \texttt{\small{read file}} test}. \diffcomment{{\proj}}{this automatically} then scaled-out to \diffcomment{74}{81} NameNodes for the 1,024 client case, \diffcomment{illustrating the efficacy of {\proj}' agile auto-scaling policy.}{This illustrates the necessity of using FaaS in the design of {\proj}, as FaaS offers auto-scaling.} 
\diffcomment{It is also worth noting that {\proj} used at-most 462.5, 425, and 475 of the 512 available vCPU during the \texttt{\small{read}}, \texttt{\small{ls}}, and \texttt{\small{stat}} client-driven scaling tests, respectively.}{It is also worth noting that {\proj} used at-most \diffcomment{475/512}{405/512} vCPUs during the client-driven scaling tests, often using closer to 300-350 vCPUs.} This illustrates {\proj}' resource efficiency, as {\proj} achieves strong performance with a fraction of the available resources.


For {\texttt{\small{create file}}} and {\texttt{\small{mkdir}}}, the performance disparity between {\proj} and HopsFS was not as significant as it was with read-based operations. The magnitude of the throughput achieved by both systems is also considerably lower than that of
read operations. Specifically, {\proj} achieved 49.09\% (1.49$\times$) higher throughput than HopsFS for {\texttt{\small{create file}}}. For {\texttt{\small{mkdir}}}, the two systems achieved roughly the same throughput. The reason both systems achieved significantly lower throughput for write operations is because the persistent metadata store quickly becomes a bottleneck. \addcomment{{\infcache} experienced poor performance for similar reasons as with the \texttt{\small{read}} operations. HopsFS+Cache also experienced low throughput, as the consistent hashing scheme used by clients can be bottle-necked by hot directories. CephFS achieved higher throughput than the other frameworks. One possible explanation for this is because CephFS' ``capabilities'' system~\cite{cephfs_capabilities} enables more efficient {\texttt{\small{write}}} operations compared to the permission system used by HopsFS and {\proj}.}

\vspace{-6pt}
\subsubsection{Resource Scaling}
\label{subsubsec:strong-scaling}
For the resource scaling experiments, the total amount of vCPUs allocated to each framework was varied between 16 and 512. 
As such, this experiment helps to elucidate how {\proj} would scale both horizontally and vertically with nearly unbounded (or at least additional) resources. Note that
for each vCPU value, all systems used the same number of clients, each of which performed \diffcomment{3,072}{2,048} operations. The largest throughput obtained is reported.

Figure~\ref{fig:strong_scale_throughput} shows the results.
For read operations (\texttt{\small{read}}, \texttt{\small{stat}}, and \texttt{\small{ls}}), {\proj} exhibited significantly better scaling than \addcomment{HopsFS, {\infcache}, and CephFS,} with 
higher throughput as the resources scaled. \addcomment{{\proj} achieved equivalent or superior throughput compared to HopsFS+Cache for all operations.}
For the largest problem size, {\proj} achieved 
\diffcomment{$30.67\times$}{$16.50\times$}, 
\diffcomment{$9.30\times$}{$5.54\times$}, and 
\diffcomment{$20.69\times$}{$4.96\times$} higher throughput than HopsFS for \texttt{\small{read}}, \texttt{\small{stat}}, and \texttt{\small{ls}}, respectively. Likewise, {\proj}' throughput increased by 
\diffcomment{$34.60\times$}{$24.24\times$}, 
\diffcomment{$34.80\times$}{$25.62\times$}, and 
\diffcomment{$72.08\times$}{$25.58\times$}
This occurred because allocating more resources to {\proj} enables a higher degree of auto-scaling. For smaller vCPU allocations, {\proj}' auto-scaling is limited and it cannot dynamically adapt to the workload, resulting in worse performance. The performance trend is less dramatic for write operations since the persistent metadata store is the bottleneck.


\begin{figure}[t]
    \centering
    \includegraphics[width=0.49\textwidth]{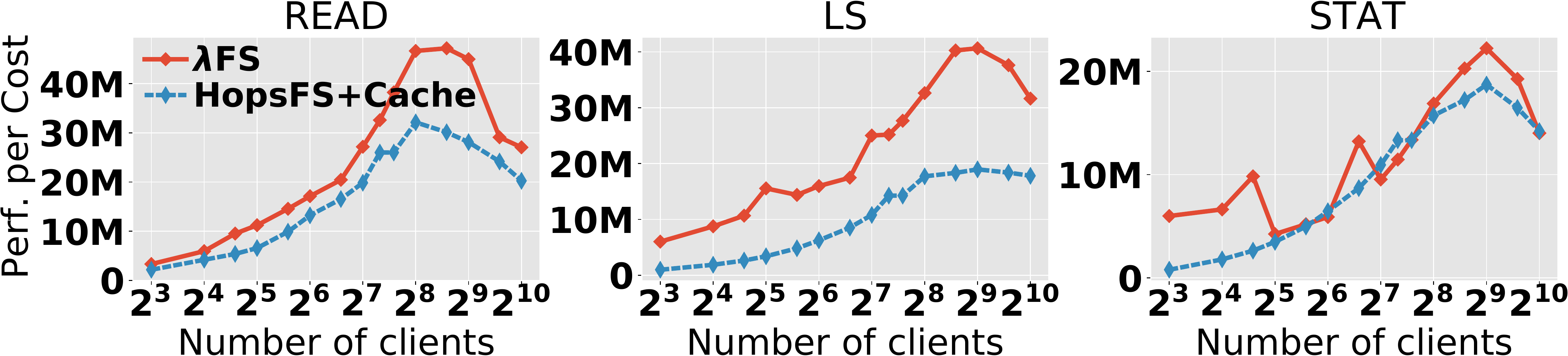}
    \vspace{-15pt}
    \caption{Performance-per-cost comparison between {\proj} and HopsFS+Cache for read-based file system operations.}
    \vspace{-10pt}
    \label{fig:perf-per-cost-client-driven}
\end{figure}

{\proj}' superior scaling behavior can once again be attributed to its metadata cache and its agile auto-scaling \diffcomment{policy}{scheme}. 
As the total amount of vCPUs increases, {\proj} provisions an increasingly large pool of concurrently-running serverless NameNodes. By using a large pool of relatively ``small'' serverless NameNodes, {\proj} achieves high resource utilization, as individual NameNodes utilize a majority of their allocated resources. This in turn enables {\proj} to achieve high performance with relatively modest resource allocations.  

Though each HopsFS NameNode was configured with 200 RPC handler threads, HopsFS was not able to fully utilize the allocated resources, because its stateless NameNodes essentially serve as proxies, forwarding requests/responses between clients and the metadata store. This also explains why HopsFS' NameNodes had a consistently low CPU utilization at around 70\%. Adding more NameNode servers may help, but again, it is difficult to pre-determine how many NameNodes to deploy for optimal performance and resource utilization.

\vspace{-2pt}
\subsubsection{Cost-Efficiency}
\label{subsubsec:cost-efficiency-microbenchmarks}

\addcomment{Figure~\ref{fig:perf-per-cost-client-driven} shows the average performance-per-cost for {\proj} and HopsFS+Cache for the \textit{client-driven} scaling tests. HopsFS+Cache's cost was computed as before: the cost of the 32 NameNode VMs running for the duration of the test. 
The cost of {\proj} was calculated using the simplified pricing model, which may have inflated the reported cost. However, since all active NameNodes were likely busy serving the high request volume for the entire experiment, the reported cost is likely close to the true cost.}

\addcomment{{\proj} achieved higher performance-per-cost values for both {\texttt{\small{read file}}} and {\texttt{\small{ls}}} for all problem sizes. For {\texttt{\small{stat file}}}, {\proj} achieved higher performance-per-cost for 8 problem sizes and roughly equivalent performance-per-cost for the others. {\proj} achieved higher performance-per-cost for {\texttt{\small{read file}}} because {\proj} achieved equivalent or higher throughput than HopsFS+Cache using a fraction of the resources; while HopsFS+Cache used 512 vCPU, {\proj} used at-most 475 vCPU by the largest problem size. This phenomenon occurred to an even greater degree for {\texttt{\small{ls}}} for which {\proj} achieved $32.74$\% higher throughput with fewer resources. For {\texttt{\small{stat file}}}, {\proj} achieved equal or better cost-effectiveness compared to HopsFS+Cache, as the two frameworks achieved similar performance, but {\proj} used fewer resources. Note that {\proj}' cost-efficiency decreased for the final few problem sizes. This occurred because {\proj} saturated an increasingly large percentage of its available 512 vCPU resources. This trend can be avoided by increasing the resources allocated to the FaaS platform, enabling {\proj} to scale-out further.}
\vspace{-6pt}
\subsection{Auto-Scaling}
\label{subsec:as}
\vspace{-2pt}

\begin{figure}[t]
    \centering
    \includegraphics[width=0.47\textwidth]{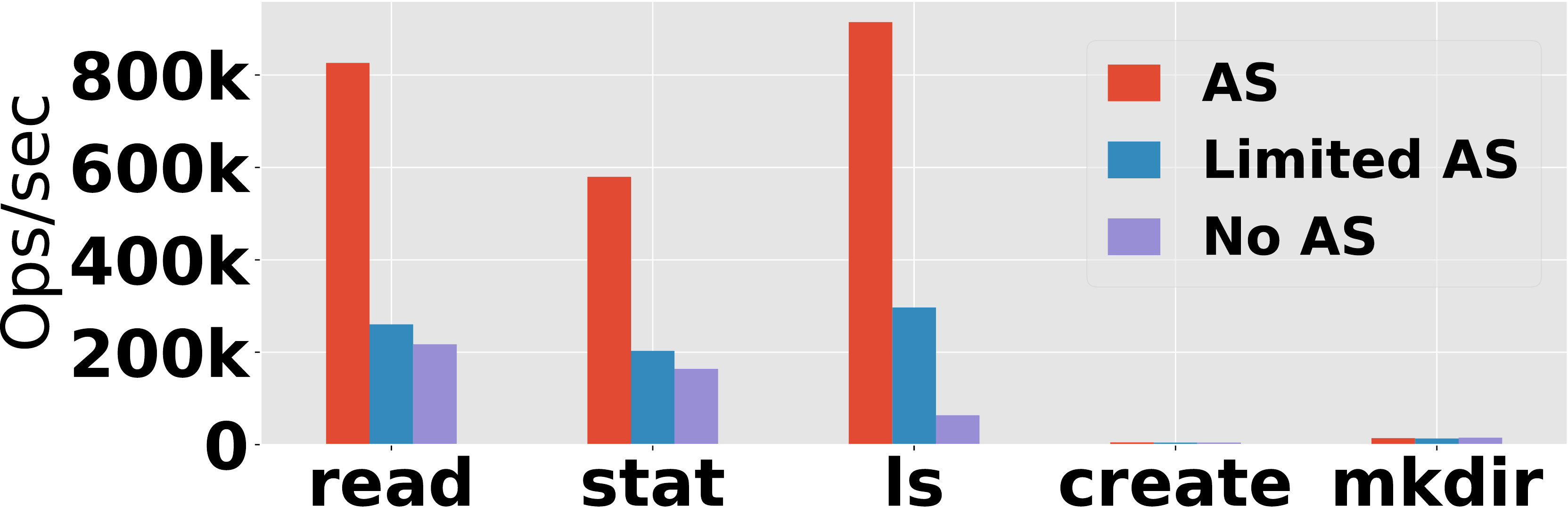}
    \vspace{-5pt}
    \caption{\diffcomment{Performance impact of auto-scaling for {\proj}}{Performance comparison for {\proj} with full, limited, and no auto-scaling}.} 
    \vspace{-5pt}
    \label{fig:auto-scaling}
\end{figure}

\addcomment{Figure~\ref{fig:auto-scaling} shows the impact on system throughput of enabling or disabling horizontal, intra-deployment auto-scaling for {\proj} across various file system operations. With auto-scaling ``enabled'', individual deployments were free to scale-out as they did in the other experiments. With limited auto-scaling, deployments could scale-out to at most 2-3 \diffcomment{active}{concurrently-running} instances. With auto-scaling disabled, each deployment was limited to a single active NameNode instance.}

\diffcomment{{\proj} achieved $2.85 - 3.17\times$ and $3.53 - 3.80\times$ higher throughput for \texttt{\small{read}} and \texttt{\small{stat file}} operations with auto-scaling enabled compared to limited and disabled auto-scaling, respectively. This trend is even more pronounced for \texttt{\small{ls}}, with {\proj} achieving $3.07\times$ and $14.37\times$ higher throughput with auto-scaling enabled compared to limited and disabled auto-scaling, respectively. The difference is less severe for write operations, as the bottleneck for writes is the persistent metadata store. These results further illustrate the importance of the FaaS-enabled agile auto-scaling policy within {\proj}' design as well as its significant impact on {\proj}' performance.}{{\proj} achieved \diffcomment{$3.17\times$}{$2.54\times$} and \diffcomment{$3.80\times$}{$4.03\times$} higher throughput for \texttt{\small{read file}} operations with auto-scaling enabled compared to limited and disabled auto-scaling, respectively. This trend continues for \texttt{\small{stat file}} operations, with auto-scaling enabling approximately \diffcomment{$2.85\times$}{$2.24\times$} and \diffcomment{$3.53\times$}{$3.81\times$} higher throughput compared to limited and disabled auto-scaling, respectively. The difference is again less severe for write operations (i.e., \texttt{\small{create file}} and \texttt{\small{mkdir}}).
This demonstrates the strong performance benefit of {\proj}' agile auto-scaling policy.}
\subsection{Subtree Operations}

\begin{wraptable}{r}{0.55\columnwidth}
\vspace{-10pt}
\centering
\centering
\caption{Average end-to-end latency (ms) of subtree {\texttt{\small{mv}}} operations for varying \texttt{\small{dir}} sizes.}
\vspace{-5pt}
\scalebox{0.9}{ 
\begin{tabular}{@{}lrl@{}}
\Xhline{1\arrayrulewidth}
\textbf{Directory Size} & \multicolumn{1}{r}{\textbf{HopsFS}} & \multicolumn{1}{r}{\textbf{{\proj}}} \\ 
\hline
262k ($2^{18}$)         & 7,511.60                            & \multicolumn{1}{r}{6,455.80}  \\
524k ($2^{19}$)         & 14,184.80                           & \multicolumn{1}{r}{12,509.20} \\ 
1.04M ($2^{20}$)        & 25,137.00                           & 25,220.80                     \\ 
\hline
\end{tabular}
} 
\vspace{-5pt}
\label{tab:subtree-mv}
\end{wraptable}

Table~\ref{tab:subtree-mv} shows the end-to-end latency of the \texttt{\small{mv}} operation performed on directories \diffcomment{whose sizes varied}{containing a large number of files. We varied the directory size} between $2^{18}$ and $2^{20}$ files. \diffcomment{On average, {\proj} completed the \texttt{\small{mv}} operation in 16.35\% and 13.39\% less time than HopsFS for $2^{18}$-file and $2^{19}$-file directories, respectively. End-to-end latency was roughly equal between HopsFS and {\proj} for $2^{20}$-file directories. For large subtree operations, the persistent metadata store becomes the bottleneck, as every write operation must update persistent state in the database.}{On average, {\proj} completed the \texttt{\small{mv}} operation in 16.35\% less time than HopsFS for the $2^{18}$-file directory. For $2^{19}$-file directories, {\proj} was about 13.39\% faster. Finally, for $2^{20}$-file directories, the two frameworks achieved roughly the same end-to-end latency. For large subtree operations, the persistent metadata store quickly becomes the bottleneck, as every write operation must update the persistent state in the database.}

\begin{figure}[t]
    \centering
    \includegraphics[width=0.47\textwidth]{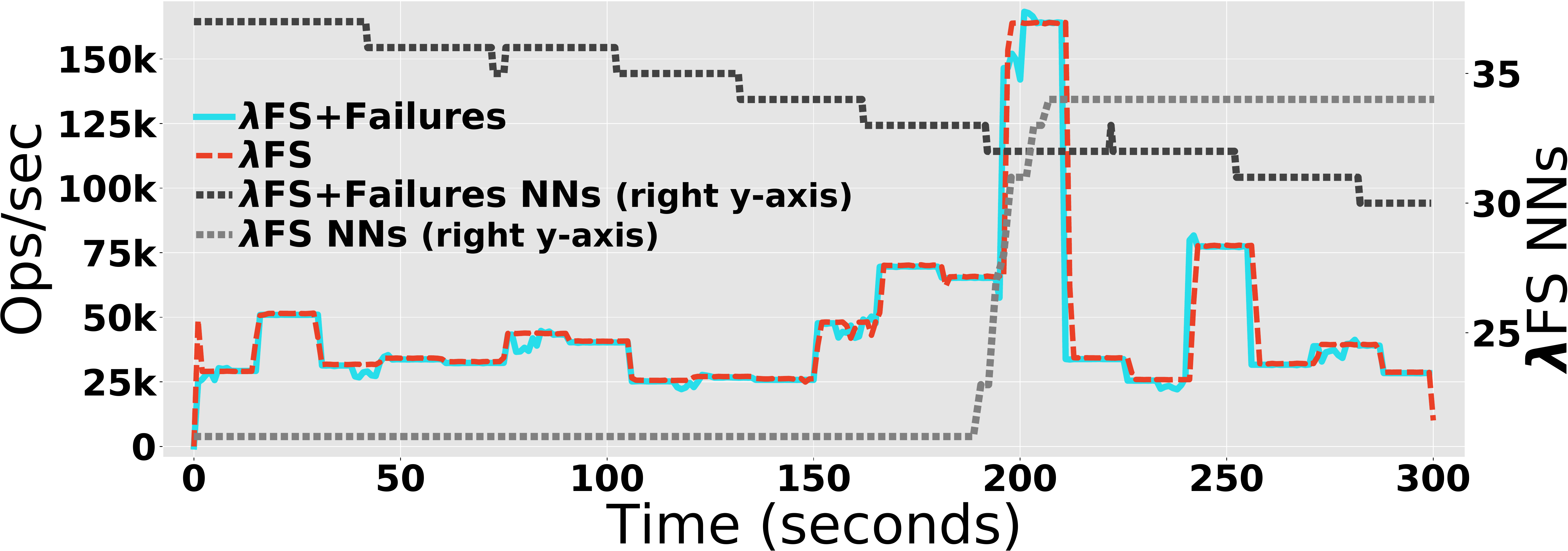}
    \vspace{-10pt}
    \caption{Fault tolerance test under the Spotify workload. 
    }
    \label{fig:spotify_25k_failure_test}
    \vspace{-10pt}
\end{figure}

\subsection{Fault Tolerance}
\label{subsec:ft}

\addcomment{To evaluate {\proj}' fault tolerance mechanisms, we executed the 25,000 ops/sec Spotify workload and manually terminated an active NameNode once every 30 seconds, targeting each deployment in a round-robin fashion. {\proj} began the workload with 36 active NameNodes (225/512 vCPU).}

\addcomment{The results of this test are shown in Figure~\ref{fig:spotify_25k_failure_test}. Despite the failures, {\proj} completed the workload as generated, even during the 163,996~ops/sec burst. The darker dashed line shows the number of active {\proj} NameNodes. {\proj}' throughput decreased slightly following a termination event, as some clients were blocked, waiting for responses to requests that had been sent to the terminated NameNode. Once these requests timed-out, they were automatically resubmitted by clients. System throughput then rose briefly as clients temporarily increased their request rate to ``make up'' for the drop in throughput that followed the termination event.}

\subsection{{\lindexfs} vs. IndexFS}
\label{sec:indexfs-results}

To \addcomment{further} demonstrate {\proj}' portability \addcomment{and performance}, we compare {\lindexfs} with IndexFS. 
%
\addcomment{For this test,} we used a 7-VM BeeGFS cluster \diffcomment{with}{that consisted of} 1 management sever, 1 metadata server, 1 storage server, and 4 \addcomment{BeeGFS} client \diffcomment{VMs}{nodes}. 
The cluster had 112 vCPUs and 448GB RAM. 
IndexFS was deployed on the 4 BeeGFS client VMs, which adheres to IndexFS' co-location principle~\cite{indexfs_sc14}. 
\lindexfs ran 1 LevelDB instance on each BeeGFS client \diffcomment{VM}{node} and used an 
OpenWhisk cluster with 
64 vCPUs and 256GB RAM to host \addcomment{the} serverless functions. 


We evaluated {\lindexfs} using IndexFS' built-in benchmarking tool {\texttt{\small{tree-test}}}.
We performed client-driven scaling \diffcomment{experiments}{tests} with the following two tests.
For the variable-sized workload, each client executed 10,000 {\texttt{\small{mknod}}} write operations followed by 10,000 random {\texttt{\small{getattr}}} read operations. For the fixed-sized workload, the total number of operations was fixed at 
1 million \diffcomment{writes}{write operations} followed by 1 million random \diffcomment{reads}{read operations}. We varied the number of clients from 2 to 256.

%

\begin{figure}[t]
\begin{center}
\subfigure[Fixed-sized workload.] {
\includegraphics[width=0.225\textwidth]{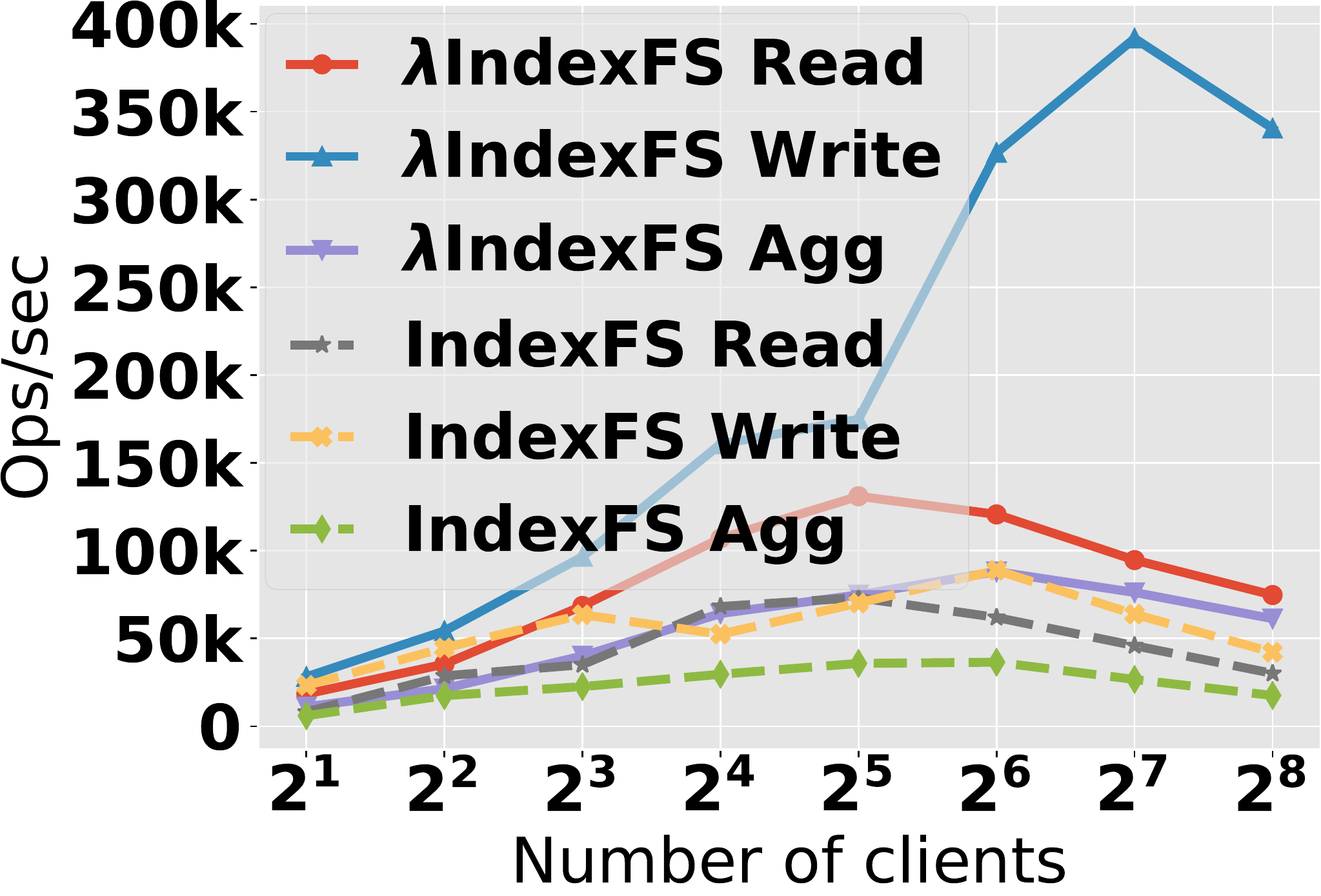}
\label{fig:indexfs-strong-scaling}
}
\hspace{-8pt}
\subfigure[Variable-sized workload.] {
\includegraphics[width=0.2125\textwidth]{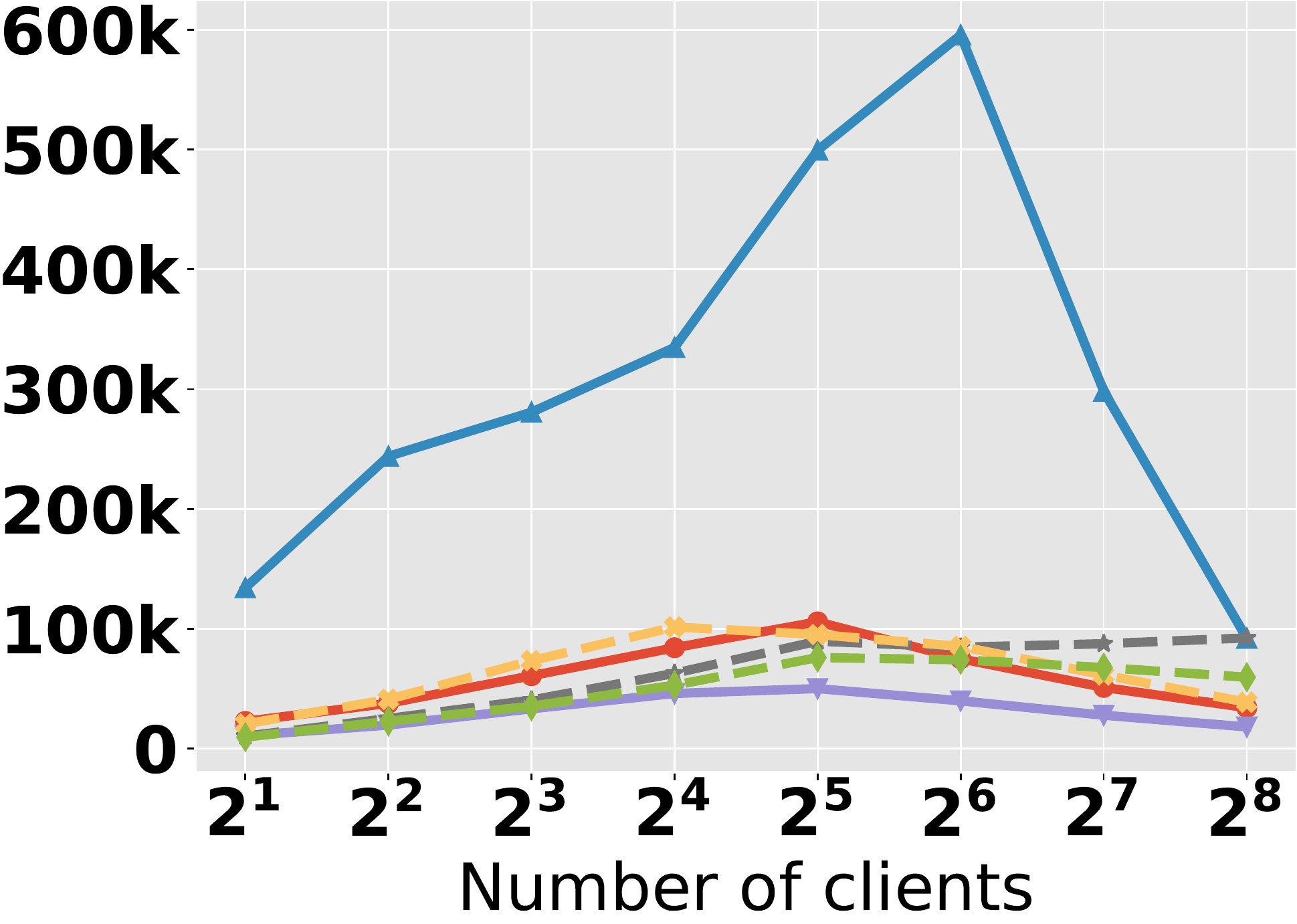}
\label{fig:indexfs-weak-scaling}
}
\vspace{-10pt}
\caption{Comparison between \lindexfs and IndexFS on BeeGFS.
{\small\texttt{Agg}} denotes the writes-followed-by-reads workload.
}
\label{fig:indexfs-scaling}
\vspace{-10pt}
\end{center}
\end{figure}

For read operations, {\lindexfs}' throughput is consistently higher than that of IndexFS, since most of the metadata is cached in serverless functions (Figure~\ref{fig:indexfs-scaling}). 
Notably, for both workloads, {\lindexfs} significantly outperforms IndexFS in terms of write throughput, largely benefiting from \diffcomment{{\lindexfs}'}{its} auto-scaling. 
{\lindexfs}' write throughput decreases when serving more than $2^6$ clients 
due to OpenWhisk's limited resources (64 vCPUs). Despite the \diffcomment{limited resources}{bottleneck}, {\lindexfs} still out-performs IndexFS, \addcomment{demonstrating the efficacy and portability of {\proj}.} 


\section{Related Work}
\label{sec:related}

\textsc{InfiniFS}~\cite{InfiniFSEfficientMetadata} implements a technique called \textit{speculative path resolution} and a client-side directory cache to optimize path resolution. 
Instead, {\proj} opts to use
HopsFS' existing ``INode Hint Cache'' to optimize path resolution. {\proj} uses \diffcomment{cloud-function-side}{serverless-function-side} caching
rather than client-side caching.

IndexFS~\cite{indexfs_sc14} is a layered, scaled-out MDS middleware built atop an existing DFS (e.g., PVFS~\cite{pvfs_als2000} and BeeGFS~\cite{beegfs}). 
IndexFS supports client-side, stateless caching, similar to the ``INode Hint Cache'' used in HopsFS and {\proj}.
{\proj} goes beyond IndexFS by offering MDS elasticity with a consistent distributed metadata cache built atop serverless functions.

LocoFS~\cite{locoFS} co-locates the metadata of a single directory on the same server, similar to how {\proj}' partitioning mechanism will co-locate metadata from single directories on the same deployments. 
This scheme can lead to single-node bottlenecks, as one metadata server can end up serving all requests for a hot directory. {\proj} avoids this bottleneck by leveraging
FaaS' auto-scaling to scale-out overloaded deployments.

Lustre~\cite{lustre} hashes on file names.
CalvinFS~\cite{thomsonCalvinFSConsistentWAN2015} hashes on full pathnames, while Giraffa~\cite{ScalingNamespaceOperations} uses full file paths as primary keys to the associated metadata. 
BetreFS uses the pathname as a file index into the local file system. 
Lazy Hybrid~\cite{lh_msst03} combines both directory subtree management and that hashing-based approach with lazy metadata relocation and lazily updated dual-entry access control lists. {\proj}
also uses hashing
but hashes on a file's parent directory. 


\textsc{InfiniCache}~\cite{infinicache_fast20} exploits the memory of AWS Lambda functions for caching large, read-only objects for low-throughput web apps. \textsc{InfiniStore}~\cite{infinistore_vldb23} is built atop \textsc{InfiniCache}, incorporating a tiered storage design that adds serverless memory elasticity and persistence. 
Faa\$T~\cite{faast_socc21} 
co-locates a key-value memory cache with a FaaS application to optimize the FaaS application's I/Os.
Pocket~\cite{pocket_osdi18} and Jiffy~\cite{jiffy_eurosys22} provide elastic, serverful, ephemeral storage for serverless analytics.
The DFS workloads that {\proj} targets are dramatically different than the applications mentioned above, therefore requiring new treatments when designing a serverless MDS system.
\section{Discussion and Lessons}
\label{sec:lessons-learned}

While designing and implementing {\proj}, we learned several interesting lessons that have applicability beyond the scope of serverless DFS metadata management. First, creating latency-sensitive applications atop FaaS requires techniques to mitigate the relatively high invocation overheads of serverless functions. Relying exclusively upon HTTP-based invocations does not enable systems to achieve high throughput and low latency; instead, mechanisms such as {\proj}' TCP-RPC invocation scheme are necessary for achieving good performance. 

Notably, introducing techniques to circumvent the large overhead of HTTP invocations can reduce the system's ability to harness the auto-scaling property of FaaS. Such techniques must be designed with care so as to effectively optimize the trade-off between maximizing performance and maximizing elasticity and scalability. This can be considered an instance of the performance-parallelism trade-off---a trade-off that has been observed in FaaS systems from other domains~\cite{wukong_socc20, wukong_pdsw19}. 

The use of FaaS also introduces a number of relatively complicated error states. Serverless functions can be reclaimed by the cloud provider at any point. If TCP-RPC connections are dropped unexpectedly, re-establishing connections is non-trivial due to the lack of addressibility of serverless functions. Additionally, naively resubmitting erred tasks via HTTP can result in request storms that overwhelm the serverless platform, leading to extreme over-provisioning of resources. This can ultimately cause a significant drop in performance and can cause errors elsewhere in the system---for example, the persistent metadata store may experience a temporary performance drop due to a wave of new connections from newly-provisioned NameNodes. To address this, FaaS-based systems must develop clever techniques to provide fault tolerance that avoids 
the aforementioned problems. 

Similarly, FaaS-based systems are intended to support hundreds or thousands of clients. If thousands of clients concurrently issue HTTP invocations, then the FaaS platform may scale-out more rapidly than is desired, quickly increasing parallelism beyond what is necessary to sustain good performance. This can lead to increased costs and thrashing-like behavior, as the system rapidly over-corrects to changes in traffic patterns. 
\section{Conclusion}
\label{sec:conclusion}

{\proj} is, to the best of our knowledge, the first cloud-native DFS metadata service, which uses the memory of serverless functions to cache and elastically scale a DFS' metadata workload. 
{\proj} achieves high-throughput, low-latency, low cost, and high resource efficiency by synthesizing a series of techniques built around a FaaS-based metadata cache. 
We have ported {\proj} to both HopsFS and IndexFS. 
We hope that this work will provide insight for building new, cloud-native, performance-sensitive backend services on FaaS.
{\proj} is open-sourced and is available at: 
\begin{center}
\url{https://github.com/ds2-lab/LambdaFS}.
\end{center}

\section*{Acknowledgments}
\label{sec:acknowledgements}

We are grateful to the anonymous ASPLOS reviewers for their valuable feedback and comments that significantly improved the paper. This work was sponsored in part by NSF grants: CNS 1943204 / 2045680 / 2322860 (NSF CAREER Awards), an NSF CloudBank grant, CCF 1910747 / 1919075 / 1919113 / 2318628, OAC-2106446, and supported by an Adobe Research gift. Benjamin Carver was supported by a Presidential Scholarship from George Mason University. 

\label{startofrefs}
\clearpage
\newpage

{
\bibliographystyle{plain}
\bibliography{main}
}

\clearpage
\appendix

\section{Straggler Mitigation}
\label{appendix:straggler-mitigation}

Tail latencies can have a detrimental impact on application performance and user experience~\cite{curtailHDFS_eurosys19,PreventingLongTail,tectonic_fast21}. 
In order to mitigate the negative impact of tail latencies, we employ a technique referred to as \textit{straggler mitigation.} {\proj} clients maintain a moving-window average latency. When a request's latency is sufficiently larger than the average (based on a configurable threshold), the request is cancelled and resubmitted to another NameNode. This can reduce the worst-case tail latencies and lead to higher system throughput. We found that the average TCP RPC latency is between 1-5ms, so we default this threshold to 10, meaning TCP requests with a latency $\geq$ 50ms will be resubmitted. 

\section{Anti-Thrashing Mode}
\label{appendix:anti-thrashing}

Typically, the FaaS platform is assumed to provide clients with virtually infinite compute and memory resources~\cite{aws_lambda}, and clients pay only for the resources they use. However, private clouds have limited cluster resources for hosting a DFS deployment~\cite{hadoop_vldb12, indexfs_sc14}. \addcomment{Additionally, to perform a fair comparison between {\proj} and HopsFS, some form of normalization is required, such as assigning equal vCPU to both frameworks, or provisioning the frameworks such that they incur the same monetary cost. However,} placing a bound on the amount of resources can result in \textit{thrashing} behavior in the FaaS platform. Recall that the serverless functions are organized into $n$ different deployments, and further that the namespace is partitioned across these deployments by hashing on the parent directory's path. Consider a scenario in which the serverless cluster's CPU utilization is approaching $100\%$. If the FaaS platform attempts to create a new container, it may have to delete an existing container to make room. When this pattern of destroying and creating containers begins to occur frequently, system throughput plummets. This is because cold starts take a non-negligible amount of time, and constantly \diffcomment{deleting}{destroying} and creating containers results in a large number of cold starts. 

To address this, client processes compute a moving average (with a configurable window size) of the latency of individual file system operations. When a metadata request observes a latency that is $T\times$ greater than the moving average latency, where $T$ is a configurable threshold parameter, the client enters \textit{anti-thrashing mode}. While in anti-thrashing mode, the client will opt to issue TCP RPCs for every metadata operation, even when no TCP connection exists to the NameNode in the deployment that is responsible for caching the requested metadata. By reusing TCP connections instead of issuing HTTP invocations, the FaaS platform will not create additional containers, as clients will issue requests to existing containers whenever possible. This will ultimately limit scaling and potentially result in reduced or leveled-off performance, but it avoids the severe performance degradation that occurs during thrashing. We find empirically that setting the threshold $T$ between 2-3 provides the best performance. 

\section{Subtree Coherence Protocol}
\label{subsec:subtree-consist-proto}

HopsFS implements subtree operations using an application-level distributed locking protocol. 
Part of this protocol involves partitioning the overall subtree operation into a number of sub-operations that are executed in-parallel.

\begin{figure}[t]
    \centering
    \includegraphics[width=.47\textwidth]{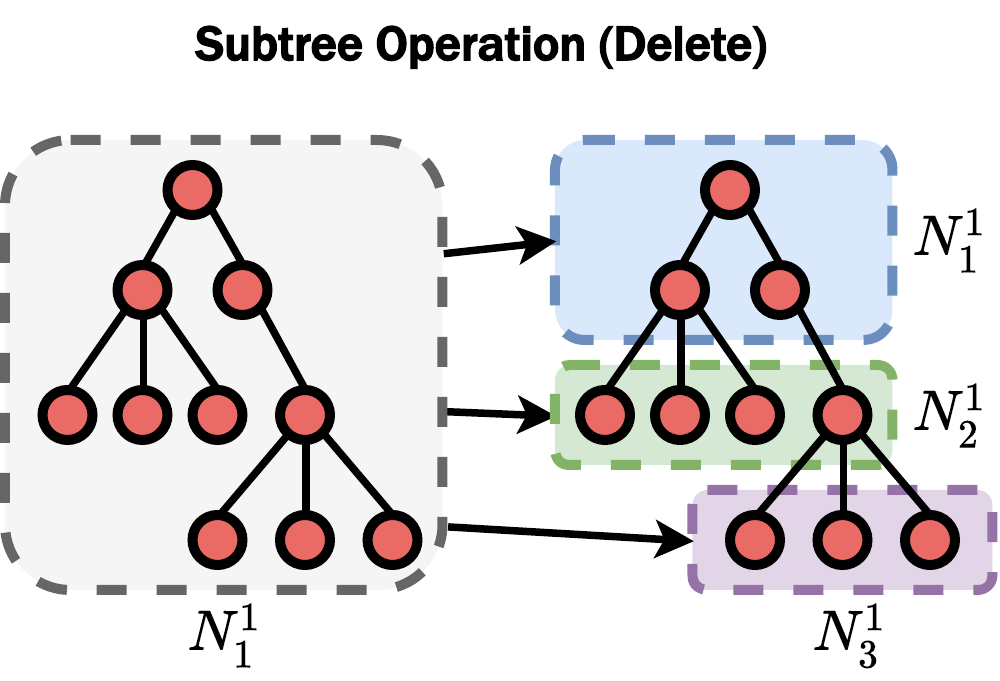}
    \vspace{-6pt}
    \caption{NameNode $N_1^1$ partitioning the sub-operations of a subtree delete operation to two other NameNodes $N_2^1$ and $N_3^1$.}
    \label{fig:subtree-op-partition-subops}
    \vspace{-5pt}
\end{figure}

There are three main phases to this protocol. In Phase~1, an exclusive lock is acquired on the subtree root, and the \textit{subtree lock} flag is persisted to the database (NDB). Active subtree operations are also stored in a table, which is queried before beginning new subtree operations in order to ensure no two operations overlap (i.e., subtree isolation). In Phase~2, the subtree is quiesced by taking and releasing database write locks on all INodes within the tree, using a predefined total ordering to avoid deadlocks. This also builds a tree data structure in-memory for use during the subtree operation. Finally, in Phase~3, the whole subtree operation is partitioned into sub-operations that can execute in-parallel. Batches of INodes are modified in each transaction in order to improve performance.

{\proj} augments the standard HopsFS subtree protocol above by integrating our serverless memory coherence protocol. A naive integration of the protocol with the standard subtree protocol would involve executing the coherence protocol once for each individual sub-operation. This would result in extremely poor performance for large subtree operations.
To address this, {\proj} performs the coherence protocol just once for the \textit{entire} subtree. This is done using a special type of invalidation, referred to as a \textit{subtree} or \textit{prefix} invalidation. Rather than specifying the individual metadata to be invalidated, {\proj} specifies the file path \textit{prefix} such that any cached INodes prefixed by this value will be invalidated. We use the subtree root as this prefix. NameNodes then utilize the trie structure of the metadata cache (\cref{subsec:design-metadata-cache}) to efficiently invalidate all INodes contained within the subtree.

Subtree invalidations are issued to all deployments responsible for caching at least one piece of metadata in the subtree. These deployments are calculated by a NameNode during a step in the Vanilla HopsFS subtree protocol. Specifically, the NameNode walks through the subtree in a predefined total order, taking out write locks. This is done to quiesce the subtree. It is during this step that we also calculate the set of deployments responsible for caching metadata in the subtree. 

As an example, consider a scenario in which the user deletes a subtree rooted at directory ``{\small\texttt{/foo/}}''. This directory may contain thousands of files and sub-directories. If the baseline coherence protocol (Algorithm~\ref{proto:consistency-protocol}) were used here, then thousands of individual invalidations would be required---one for each INode
within the subtree. Instead, the leader NameNode simply issues a single \textit{subtree} invalidation with the prefix ``{\small\texttt{/foo/}}'' to all deployments caching any metadata within the subtree. Once the leader NameNode has received all the {\small\texttt{ACKs}}, {\proj} is free to execute the subtree operation without running any further instances of the coherence protocol.

\noindent\textbf{Elastically Offloading Batched Operations.} 
The sub-operations created during subtree operations are typically executed in-parallel on the NameNode orchestrating the subtree operation. This works well when each NameNode has a large amount of CPU resources allocated to it. Serverless NameNodes, however, typically have a small amount of CPU cores allocated. As a result, executing hundreds or thousands of operations can be slow. 
To address this, we designed a technique referred to as serverless offloading. That is, {\proj}
offloads batches of sub-operations to other NameNodes
by taking advantage of FaaS elasticity in order to increase parallelism and scalability. The overhead of the coherence protocol is therefore minimized, thanks to batching and serverless offloading.

The batch size is a configurable parameter. We found that larger batch sizes tend to perform better, as there is a trade-off between increasing parallelism and the network overhead of offloading the operations. The batch size parameter defaults to 512. 
Figure~\ref{fig:subtree-op-partition-subops} illustrates an example of this procedure. In Figure~\ref{fig:subtree-op-partition-subops}, we refer to the $i^{th}$ NameNode in the $j^{th}$ deployment as $N_i^j$. We say that a NameNode $N$ belongs to deployment $D_i$ (the $i^{th}$ deployment) using $N \in D_i$.

In this example, the client sends a ``{\small\texttt{rm -rf /foo/bar}}'' operation to NameNode $N_1^1$, which caches all of the files and sub-directories rooted under {\small\texttt{/foo/bar}}; $N_1^1$ offloads level 2 and level 3 of the subtree to a different set of helper NameNodes, $N_2^1$ and $N_3^1$, from deployment 2 and 3. This does not create a consistency problem as the helper NameNodes simply help $N_1^1$ process part of $N_1^1$'s load to speedup the subtree processing. 

\end{document}